\documentclass[journal]{IEEEtran}
\usepackage{multirow}
\usepackage{float}
\usepackage[dvips]{graphicx}
\usepackage{amsfonts}
\usepackage{amsmath}
\usepackage[psamsfonts]{amssymb}
\usepackage{amsxtra}
\usepackage{threeparttable}
\usepackage{bezier,amsmath,amsfonts,amssymb,epsfig,cite,bm,setspace,color,url,balance}
\usepackage{tabulary}
\usepackage{algpseudocode}
\usepackage{color}
\usepackage{amsthm}
\usepackage{cite}
\usepackage{graphicx}
\usepackage{epstopdf}
\usepackage{stfloats}

\usepackage{color}
\usepackage{bm}
\usepackage[linesnumbered,boxed,ruled,commentsnumbered]{algorithm2e}

\usepackage{booktabs}
\usepackage{float}
\usepackage{cite}
\usepackage{graphicx}
\usepackage{epstopdf}
\usepackage{subfigure}
\usepackage{algpseudocode}
\usepackage{graphicx}
\usepackage{float}
\usepackage{color}
\usepackage{subfig}
\usepackage[psamsfonts]{amssymb}
\usepackage{amsxtra}
\usepackage{threeparttable}
\usepackage{bezier,amsmath,amsfonts,amssymb,epsfig,cite,bm,setspace,color,url,balance}
\usepackage{tabulary}
\usepackage{amsthm}
\usepackage{makecell}
\usepackage{booktabs}
\usepackage{mathtools}
\usepackage{algpseudocode}
\usepackage{mathrsfs}
\usepackage{bbm}
\usepackage{balance}

\begin{document}
		\title{Next-Generation Full Duplex Networking System Empowered by Reconfigurable Intelligent Surfaces}
		\allowdisplaybreaks
			 \author{Yingyang Chen,~Yuncong Li,~Miaowen Wen,~Duoying Zhang,\\ Bingli Jiao,~Zhiguo Ding,~Theodoros A. Tsiftsis,~and  H. Vincent Poor
	 	
	 	\thanks{Y. Chen, Y. Li and D. Zhang are with the Department of Electronic Engineering, College of Information Science and Technology, Jinan University, Guangzhou 510632, China (e-mail: \{chenyy; JingcongLi; zhangduoying\} @jnu.edu.cn).}
	 	
	 	\thanks{M. Wen is with the School of Electronic and Information Engineering, South China University of Technology, Guangzhou 510640, China (e-mail: eemwwen@scut.edu.cn).}

	 	\thanks{B. Jiao is with School of Electronics Engineering and
	 		Computer Science, Peking University, Beijing (e-mail: jiaobl@pku.edu.cn).}

	 	\thanks{Z. Ding is with the School of Electrical and Electronic Engineering, The University of Manchester, Manchester M13 9PL, U.K. (e-mail: zhiguo.ding@manchester.ac.uk).}	
	 	
	 	\thanks{T. A. Tsiftsis is with the School of Intelligent Systems Science and
	 		Engineering, Jinan University, Zhuhai 519070, China. (e-mail: theotsiftsis@jnu.edu.cn).}
	 	
	 	\thanks{H. V. Poor is with the Department of Electrical and Computer Engineering, Princeton University, Princeton, NJ 08544 USA (e-mail: poor@princeton.edu).}
	 	
	 }
		\maketitle	
		
		\begin{abstract} 
			
			Full duplex (FD) radio has attracted extensive attention due to its co-time and co-frequency transceiving capability. {However, the potential gain brought by FD radios is closely related to the management of self-interference (SI), which imposes high or even stringent requirements on SI cancellation (SIC) techniques. When the FD deployment evolves into next-generation mobile networking, the SI problem becomes more complicated, significantly limiting its potential gains.} In this paper, we conceive a multi-cell FD networking scheme by deploying a reconfigurable intelligent surface (RIS) at the cell boundary to configure the radio environment proactively. To achieve the full potential of the system, we aim to maximize the sum rate (SR) of multiple cells by jointly optimizing the transmit precoding (TPC) matrices at FD base stations (BSs) and users and the phase shift matrix at RIS. Since the original problem is non-convex, we reformulate and decouple it into a pair of subproblems by utilizing the relationship between the SR and minimum mean square error (MMSE). The optimal solutions of TPC matrices are obtained in closed form, while both complex circle manifold (CCM) and successive convex approximation (SCA) based algorithms are developed to resolve the phase shift matrix suboptimally. Our simulation results show that introducing an RIS into an FD networking system not only improves the overall SR significantly but also enhances the cell edge performance prominently. More importantly, we validate that the RIS deployment with optimized phase shifts can reduce the requirement for SIC and the number of BS antennas, which further reduces the hardware cost and power consumption, especially with a sufficient number of reflecting elements. As a result, the utilization of an RIS enables the originally cumbersome FD networking system to become efficient and practical.
			
		\end{abstract}
		
		\section{Introduction}
		   
 \subsection{Motivation}       
 \IEEEPARstart{T}{he} ongoing explosive growth of data-hungry applications puts forward ultimate requirements for wireless communications. However, with limited spectrum resources, improving the communication rate through legacy half-duplex (HD) communications is quite difficult. To address this limitation,  the community has turned its attention to several emerging technologies, such as millimeter wave (mmWave) communications, massive multiple-input multiple-output (MIMO),  and full duplex (FD) communications \cite{FullDuplexMultiUserMultiCell}. Among these, considering the appealing spectral efficiency (SE) and the co-time co-frequency transceiving capability compared to its HD counterpart, FD has attracted significant attention in the hope of achieving its full potential.
        
        Different from the HD mode, in FD communications, two or more terminals can exchange information over the same frequency band simultaneously. Therefore,  in theory, FD radio has the potential to double the SE \cite{FullDuplexWirelessCommunications, DynamicSpectrumSharing}. However, the mechanism of allocating transmit and receive signals in co-frequency and co-time inevitably introduces serious interference problems. Among them, the most impactful one is the self-interference (SI) from transmit to receive antennas, which is far stronger than the received signal. The uplink (UL) rate will suffer quite a loss if the SI is not eliminated sufficiently. Consequently, significant research efforts have been devoted to SI cancellation (SIC) techniques \cite{Fullduplextechniquesfor5G,Applicationsofselfinterferencecancellation}, which advanced the practicability of FD deployment, especially in single-cell communications \cite{Fullduplexdevicetodevice,fullduplexsmallcellwireless}.
        
        Nevertheless, when the mobile deployment evolves into next-generation FD networking, the receivers will suffer from more additional interference, not only from their own transmitters, but also from the neighboring cells, which greatly limits the potential gains of using FD technology \cite{FDcellualr}. To cope with the complex radio environment in networking, active beamforming with antenna arrays is often proposed to apply a directional beam and align the signal to the intended receiver \cite{FD_ActiveBeamforming1}. {However, to further improve the system performance, more antennas and radio frequency (RF) chains are needed, requiring sophisticated signal processing techniques and further resulting in lower energy efficiency (EE) and higher hardware cost. {Although hybrid beamforming can be applied to reduce the RF chain cost, this still incurs high complexity to avoid performance erosion.}} Therefore, the application of FD radios in wireless networking remains problematic due to the aforementioned issues.
        
        With the revolution in micro-electrical-mechanical system and programmable meta-materials, reconfigurable intelligent surfaces (RISs) have latterly received extensive attention as they can improve both SE and EE at low cost \cite{Smartradioenvironments}. Specifically, an RIS comprises an array of nearly passive reflecting elements, each of which can independently apply a phase shift to an incident signal, and thus a favorable channel can be designed for wireless transmission. By carefully adjusting the phase shift matrix of an RIS, the reflected link can be superimposed constructively at the desired receiver or added deconstructively at unintended terminals. Hence, some recent studies have incorporated RISs into wireless communication systems to eliminate serious interference and demonstrated its effectiveness in applications such as FD bi-directional communication systems \cite{MultiuserFullDuplexTwoWayCommunications}, unmanned aerial vehicles (UAV) aided communications \cite{UAVCommunicationWithRIS}, integrated sensing and communications (ISAC)  \cite{IntegratedSensingandCommunication}, and wireless edge caching \cite{ChenTWC}. Thus, RIS technology has the potential to provide efficient solutions for interference limited systems by shaping wireless propagation environments proactively, which may serve as a feasible  way to improve FD networking systems. To reap the benefits of using an RIS, the phase shift matrix of the RIS should be properly optimized along with the active beamforming at FD transceivers.

		\subsection{Related Works}
		\subsubsection{FD in Wireless Networks}
		 Because of the leaky signal from the transmit antennas to the receive ones, SIC is a pivotal technique in FD deployment \cite{IntegratedAccessandBackhaul,DesignAndAnalysis,JointAnalog,HybridBeamforming}. In \cite{IntegratedAccessandBackhaul} the authors reviewed the existing SI reduction techniques, and validated the attenuation level through hardware prototype. An optical domain (OD)-based analog SIC was studied in \cite{DesignAndAnalysis} to reduce both the space and cost. In \cite{JointAnalog},  a truncated singular value decomposition (TSVD)-based digital SIC method was explored by Islam \textit{et al.}, focusing on the leading terms of residual SI, thus reducing the estimated parameters. In \cite{HybridBeamforming}, the authors introduced a hybrid beamforming method to prevent residual SI signals from saturating under hardware conditions. Inspired by the substantial SE gain brought by FD radios, there exist a large number of works investigating the interaction between FD with other emerging technologies, e.g.,  UAVs \cite{MillimeterWave},  wireless edge caching \cite{Chen20TNSE}, 
		backscatter communication \cite{MaximizingSecondary}, wireless powered communication network (WPCN) \cite{MultiDomain}, and ISAC \cite{WaveformDesign}. {Specifically, Zhu \emph{et al.} \cite{MillimeterWave} proposed to deploy an FD-UAV relay with large antenna arrays to enhance the mmWave channels.
		In \cite{Chen20TNSE}, FD relaying was merged into the content hitting and retrieving procedures. The derived successful probability validated that FD benefits the wireless caching systems under a certain degree of SIC capability. Jafari \emph{et al.} \cite{MaximizingSecondary} investigated the optimal time and energy allocation to maximize the sum throughput in an FD cognitive backscatter communication network. In \cite{MultiDomain}, the authors studied transmit beamforming and receive combing at the FD access point, as well as user equipments (UEs) scheduling in WPCN to overcome the doubly near-far effect. Furthermore,  Xiao \emph{et al.} \cite{WaveformDesign} applied FD radios to ISAC by fully utilizing the waiting time of the conventional pulse radar to transmit the communication signal. It was validated that, when the SIC is greater than a threshold, the communication and sensing performance are both improved. \textbf{From the above mentioned works, we deduce that the potential gain brought by FD radios is closely related to the SI management, which imposes high or even severe requirements on SIC capabilities.}}
		
		In addition to SI, there also exists other interference brought by FD radios waiting to be tackled. Typically, there are a number of works that study the FD operation in a single cell (see \cite{Fullduplexdevicetodevice,FullDuplexMIMOInterferenceChannels,fullduplexmassiveMIMOsystems}). {When the FD technique is employed in a more practical deployment, typically in a multi-cell multi-user network, it would face a much more complicated radio environment. Explicitly, the inter-cell interference becomes prominent in an FD cellular network. To cope with this issue, many prior works have been carried out for exploration \cite{LowComplexity,FullDuplexMultiUserMultiCell,LinearTransceiverDesignforFullDuplex,FD_ActiveBeamforming1,Ming_min_Zhao,Wang_FD_Hybrid,Fullduplexcellularsystems}.} {In \cite{FullDuplexMultiUserMultiCell}, the authors designed the beamforming at FD BSs and HD users (evolve to FD terminals in \cite{LinearTransceiverDesignforFullDuplex}) to suppress both intra-cell and inter-cell interference in a multi-small-cell network. Bai \emph{et al.} \cite{FD_ActiveBeamforming1} characterized the ergodic rate performance for the multi-cell FD network, and showed the double gain brought by FD when the number of BS antennas goes infinite. Zou \emph{et al.} \cite{LowComplexity} proposed coordinated beamforming at FD BSs to cope with BS-BS interference, while UE-UE interference  was suppressed by users' scheduling. \textbf{However, beamforming armed with large antenna arrays brings better interference management at a price of complicated structure and high power consumption as well as hardware cost.}} Alternatively, the authors in \cite{Ming_min_Zhao} proposed a robust hybrid beamforming design for an FD mmWave multi-cell system. However, it assumed perfect SIC, and required the phase shifter with qualified quantization resolution. The authors in \cite{Wang_FD_Hybrid} proposed a two-step hybrid beamforming method to further eliminate the residual SI and compensate for the SE loss, while it required all transceivers upgrade into hybrid mode. Ma \emph{et al.} \cite{Fullduplexcellularsystems} developed a prototype of the FD networking system that applies distributed transmit and receive antenna arrays at each FD BS to suppress the SI and BS-BS interference. However, such bistatic deployment increases the hardware cost, and the proposed two-step beamforming design achieves limited performance.
		  \textbf{How to manage the interference environment efficiently in FD networking systems still remains to be explored.}

		\subsubsection{RIS-Empowered Wireless Networks}	
			
		From the perspective of RIS, quantities of work have been devoted to getting the utmost out of it \cite{CascadedChannel,DeepLearning, RISAidedWirelessCommunication,WirelessCoverageExtension,ActiveReconfigurableIntelligentSurface,MultipleReconfigurableIntelligentSurfaces}. Due to the lack of signal processing capability and abundant reflecting elements on RIS, efficient channel estimation methods are required. Particularly, in \cite{CascadedChannel}, one reference user with the FD functionality was selected as an anchor point to estimate the common RIS-BS channel, hence reducing the required training overhead. Zhang \textit{et al.} \cite{DeepLearning} assumed that several elements as a group share an identical channel and utilized deep learning for estimation, further lessening the overhead. Besides, the studies of \cite{RISAidedWirelessCommunication} and \cite{WirelessCoverageExtension} explored the optimal orientation and location of RIS to earn extra degrees of freedom and coverage extension, respectively. On account of the twice large-scale fading in the RIS-aided link, the authors in \cite{ActiveReconfigurableIntelligentSurface} proposed an active RIS model, which is capable of amplifying the reflected incident signal to resist the double path loss. Meanwhile, a group of researches focused on integrating RIS into other emerging topics \cite{InterceptProbability,UAVCommunicationWithRIS,IntegratedSensingandCommunication,aReconfigurableIntelligentSurfaceandaRelay,RobustBeamforming,HardwareImpairments}. The authors in \cite{UAVCommunicationWithRIS} put forward an RIS-aided FD UAV communication, using RIS to mitigate the co-channel interference from the UL user, while regarding the residual SI as noise. Wang \textit{et al.} \cite{IntegratedSensingandCommunication} designed an RIS-assisted ISAC system to minimize the multi-user interference (MUI). Zaghdoud \textit{et al.} \cite{InterceptProbability} proposed to heighten the jamming signal transmitted by the FD device and the legitimate signal for the intended receiver with the assistance of RIS.  To reap the benefits of both RIS and cooperative relay, the authors in \cite{aReconfigurableIntelligentSurfaceandaRelay} proposed to apply them in a multi-user multiple-input single-output (MISO) downlink (DL) system. It proved that RIS could play its role more effectively with the existence of a relay. The robustness of the RIS aided system was also studied under imperfect channel state information (CSI) and hardware impairments in \cite{RobustBeamforming} and \cite{HardwareImpairments}, respectively.
		
		Moreover, significant contributions have been devoted to empowering the legacy wireless communication networks with RIS \cite{OptimalControlforFullDuplex,SumRateOptimization,WeightedSumRateMaximization,MultiuserFullDuplexTwoWayCommunications,MultipleReconfigurableIntelligentSurfaces,MulticellMIMOCommunications}. A point-to-point RIS-aided communication system with only a reflected link was studied in \cite{OptimalControlforFullDuplex}, where two devices communicate in FD mode. It showed that the proposed scheme could reduce up to two-thirds of the transmit power compared to that operating in HD mode when SIC is powerful enough. Further, the authors in \cite{SumRateOptimization} considered direct and reflected links, where both continuous and discrete phase shifts are optimized. Guo \textit{et al.} \cite{WeightedSumRateMaximization} maximized the weighted sum rate (SR) of an RIS-aided multi-user MISO DL communication system. A multi-user FD two-way communication system was explored in \cite{MultiuserFullDuplexTwoWayCommunications}, where the formulated non-convex problem is first decoupled and then solved by the block coordinate descent (BCD) and minorization-maximization (MM) algorithms. However, the residual noise resulting from the interference cancellation at FD transceivers was simplified as pure additive white Gaussian noise (AWGN), which fails to reflect the impact of SIC capability. Nguyen \textit{et al.} \cite{MultipleReconfigurableIntelligentSurfaces} studied a bidirectional FD communication system with multiple RISs assisted. Their analytical outage probability and ergodic capacity results showed that the impact of SI can be reduced by enlarging the size of RISs.
		{Unlike the aforementioned works considering one single cell, the study of \cite{MulticellMIMOCommunications} designed a multi-cell RIS-aided HD communication network, which takes intra- and inter-cell interference into full consideration for the DL data transmission. However, the UL rate will suffer quite a loss due to the severe BS-BS interference.} \textbf{At the time of writing, there is a paucity of results on the utilization of RIS to the sophisticated multi-cell FD networking system.}
		  
	    \subsection{Contributions and Organizations}
	    Against this background, our main contributions in this paper are three-fold:	    
	    \begin{itemize}
	    	\item Firstly, this is an early exploration of empowering a multi-cell FD networking system with the assistance of RIS. The problem of maximizing the SR of all cells is formulated to optimize both active beamforming at FD BSs and HD users and passive beamforming at RIS,  subject to the power budget and unit modulus constraints. The formulated problem cannot be solved straightforwardly, since the transmit precoding (TPC) matrices and the phase shift matrix are highly coupled.
	    	
	    	\item  Secondly, we decouple the original objective function (OF) by exploiting the relationship between SR and minimum mean square error (MMSE), and then optimize the TPC matrices and phase shifts in a BCD manner. Given the fixed phase shift matrix, the optimal solutions for TPC matrices are derived in closed form through Lagrangian multiplier method and the bisection search. For the phase shift optimization, we develop a complex circle manifold (CCM) and a successive convex approximation (SCA)-based method to obtain the suboptimal solution, respectively, where the latter is validated to have low complexity.
	    	
	    	\item  Finally, we present numerical validations and evaluations.  It is shown that the RIS-empowered FD scheme could achieve the same performance with a much moderate SIC capability as the scheme without RIS whilst requiring stringent SIC. Meanwhile, the proposed scheme achieves prominent cell-edge performance enhancement and can decrease the antenna array size at BS transceivers, especially with a sufficient number of reflecting elements. All these observations validate that deploying RIS can promote FD radios to evolve into mobile networking efficiently.
	    
	    \end{itemize}

Again, instead of simply applying RIS to FD radios, we intrinsically harness the potential provided by RIS to promote the practical deployment of FD networking. The rest of this treatise is organized as follows. Section II presents the system model and optimization problem for RIS empowered FD networking. Next, Section III discusses the design of TPC matrices at FD BSs, while Section IV provides a pair of algorithms to optimize the phase shifts. Simulation results and discussions are provided in Section V. Finally, Section VI draws a conclusion. The most frequent notations are listed here for convenience.
    
  	\begin{figure}[t]
  	\centering
  	\includegraphics[scale=0.25]{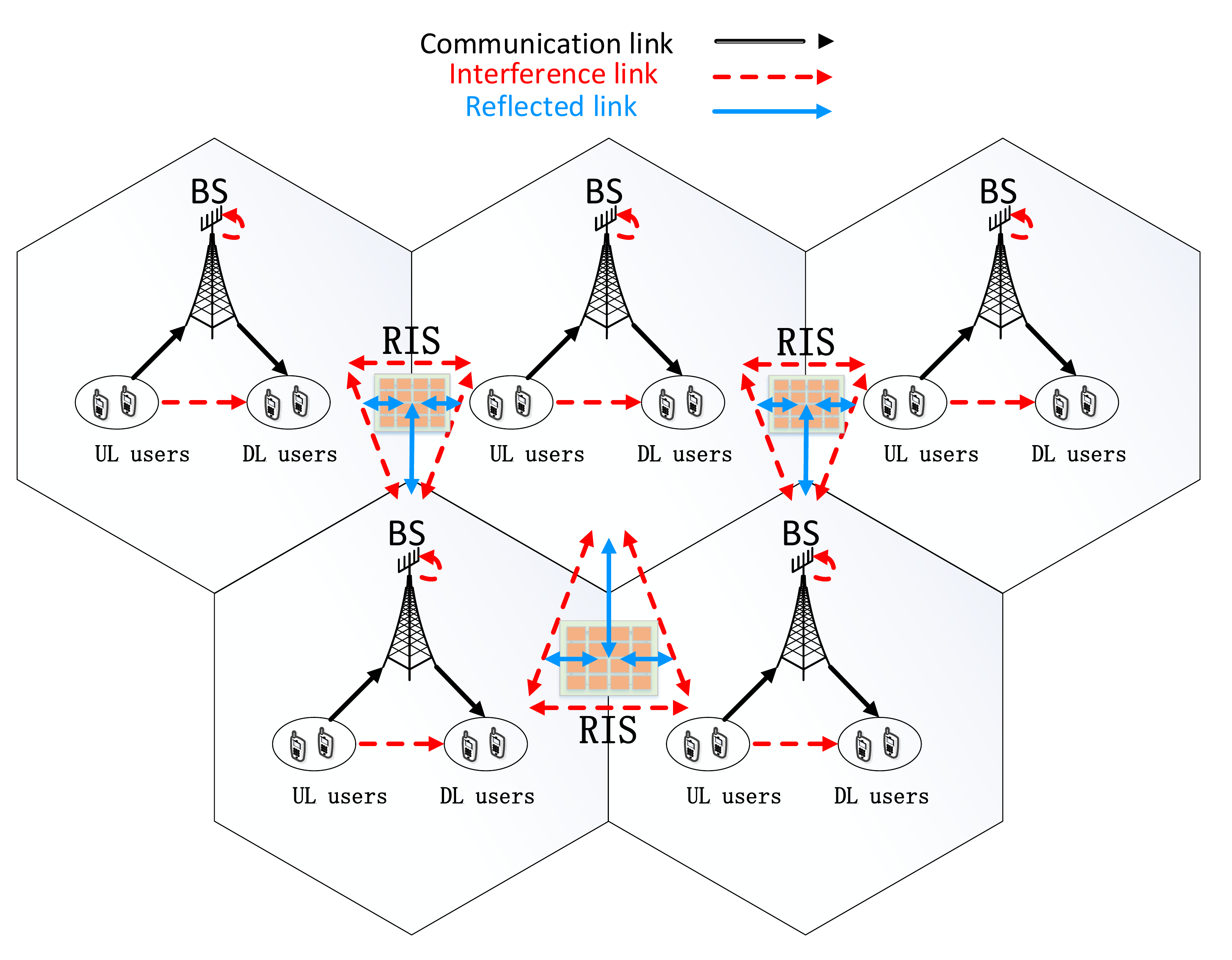}
  	\caption{The proposed RIS-aided multi-user multi-cell FD mobile networking system.}
  	\label{fig:system model1}
  \end{figure}		
                            
        \textit{Notations:} Uppercase and lowercase bold-faced letters indicate matrices and vectors, respectively. ${\mathbf{A}} \in {\mathbb{C}^{M \times N}}$ denotes  a complex-element matrix with dimensions ${M \times N}$. ${{\mathbf{I}}_N}$ represents the $N \times N$ dimensional identity matrix. Let ${{\mathbf{A}}^H}$, ${{\mathbf{A}}^T}$ and ${{\mathbf{A}}^*}$ refer to the Hermitian, transpose and conjugate of the matrix ${\mathbf{A}}$, respectively. The Hadamard product is denoted by $ \odot $. ${\rm T}{\text{r}}\left(  \cdot  \right)$ and $\det \left(  \cdot  \right)$  are the trace and determinant operators of a complex argument, respectively. Symbol ${\text{diag(}} \cdot {\text{)}}$ denotes the diagonalization operation. The  expectation and real part of a complex matrix are denoted by  $\mathbb{E}\left[ {\mathbf{A}} \right]$ and ${\text{Re(}}{\mathbf{A}}{\text{)}}$, respectively.  For a complex scalar, $\left| a \right|$ denotes the absolute value of $a$. For a vector, the Euclidean norm is denoted as ${\left\| {\mathbf{a}} \right\|_2}$. For a matrix, ${\left[ {\bf{A}} \right]_{i,i}}$  denote the $i$th diagonal element of matrix ${\bf{A}}$. Finally, $x \sim {\mathcal {CN}}\left( {\mu ,{\sigma ^2}} \right)$ indicates that the random variable $x$ obeys a complex Gaussian distribution with mean $\mu$ and variance $\sigma^2$.


		\section{System Model and Problem Formulation}
		\subsection{System Model}

	\begin{figure}[t]
	\centering
	\includegraphics[scale=0.23]{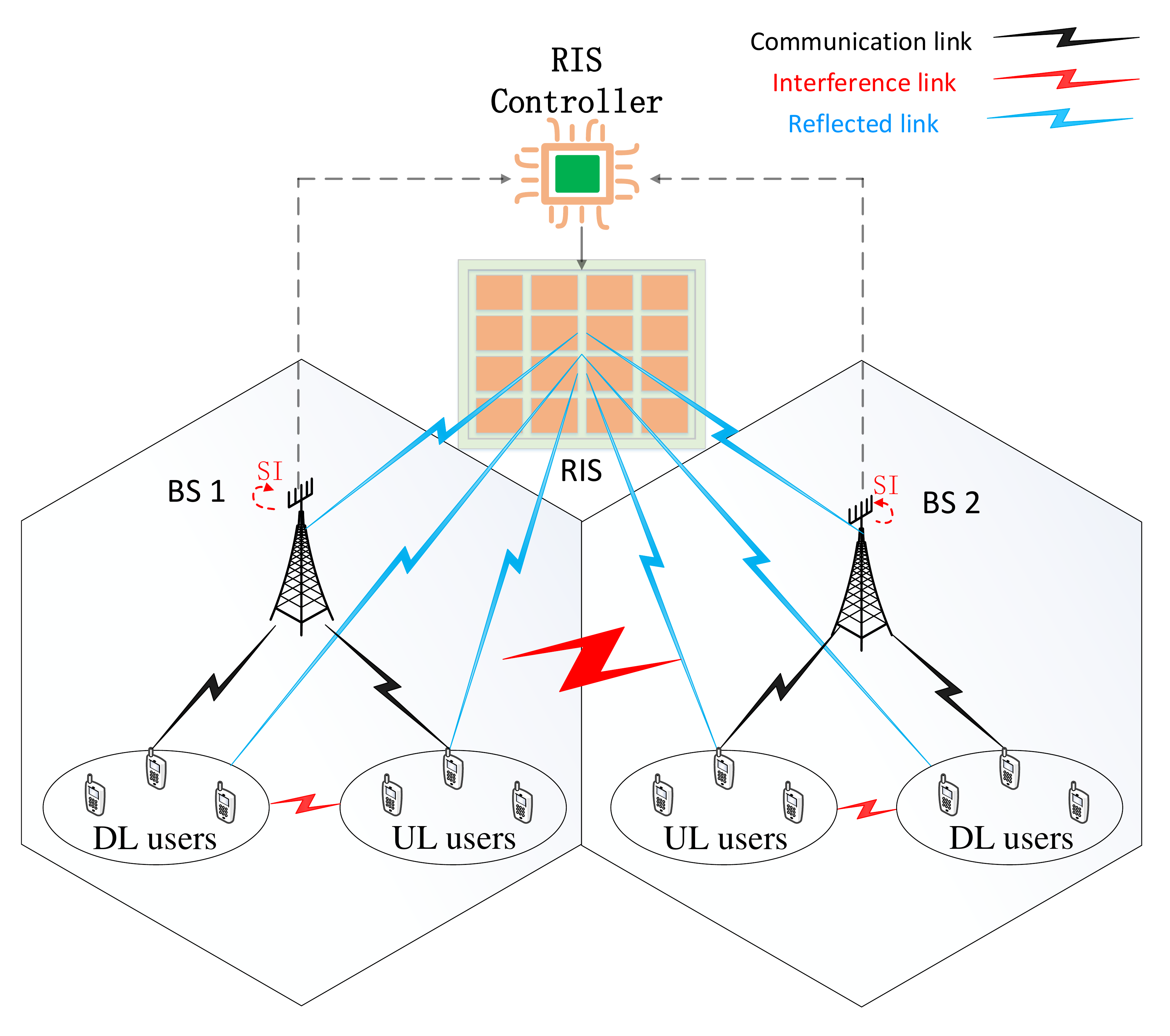}
	\caption{An illustration of the RIS-aided FD communication system between two cells.}
	\label{fig:system model}
\end{figure}

		{As shown in Fig. \ref{fig:system model1}, we conceive a multi-user multi-cell FD networking system empowered by RISs, where each cell includes one BS, multiple UL and DL users. The BS and users are assumed to operate in the FD and HD modes, respectively. On the edge of cells, we deploy RISs with sufficient passive reflecting elements. The seminal idea of the conceived system is to configure the phase shifts at RISs and precoding at BSs and user transceivers to provide a favorable wireless environment for transmissions. Particularly, we assume that  in the networking system, each BS is equipped with $\mathop M\nolimits_B^t $ transmit and $\mathop M\nolimits_B^r $ receive antennas, while all users are  equipped with $\mathop M\nolimits_U^t $ transmit and $\mathop M\nolimits_U^r $ receive antennas, respectively. Each RIS has ${M}$ passive reflecting elements. Clearly, the radio environment of the considered system is complicated, due to the coexistence of multiple direct and reflected links as well as various intra- and inter-cell interference. }
			
		To have a close look, Fig. \ref{fig:system model} depicts a two-cells FD system with one RIS deployed.\footnote{In the following, we investigate the system with multiple FD cells and one RIS plane, in the hope of gaining insights on the feasibility firstly. The overall design for the scheme with multiple RIS planes would be left for future discussions.} To improve the receiving performance, we aim to adjust the phase shift matrix at RIS and the TPC matrices at BSs and users jointly to configure the wireless environments proactively.  Specifically, we consider there are ${{L}}$ cells, where each cell has $\mathop K\nolimits_{}^d  $ DL users and $\mathop K\nolimits_{}^u $ UL users. For simplicity, we use $d$ and $u$ as superscript to indicate the DL and UL users, respectively, and denote a set $\mathcal{V} \buildrel \Delta \over =\left\{ {u,d} \right\}$. Hence ${k_l^v}$ implies the $k{\text{th}}$ DL or UL user in the $l{\text{th}}$ cell, where $v\in\mathcal{V}$ and $l \in \left\{ {1, \cdots ,L} \right\} \buildrel \Delta \over =  \mathcal{L}$.	Specifically, the signal transmitted to the $ {\mathop k\nolimits_l^d }$th DL user from the $l$th BS is given by
		\begin{equation}
			\mathop {\bf{x}}\nolimits_{\mathop k\nolimits_l^d }  = \mathop {\bf{F}}\nolimits_{\mathop k\nolimits_l^d } \mathop {\bf{s}}\nolimits_{\mathop k\nolimits_l^d } ,  
		\end{equation}\label{}\noindent 
		where ${\text{ }}\mathop {\mathbf{s}}\nolimits_{\mathop k\nolimits_l^d }  \in {\mathbb{C}^{\mathop b\nolimits_d  \times 1}}  $ denotes the ($\mathop b\nolimits_d $)-dimensional symbol vector with unit power transmitted by the $l$th BS to the $ {\mathop k\nolimits_l^d }$th DL user, and $\mathop {\bf{F}}\nolimits_{\mathop k\nolimits_l^d }  \in {\mathbb{C}^{\mathop M\nolimits_B^t  \times \mathop b\nolimits_d }}$ is the corresponding TPC matrix, meeting the constraint of the maximum power budget $\mathop P\nolimits_B$ at the BS side as follow
		\begin{equation}
			\sum\limits_{k = 1}^{\mathop K\nolimits_l^d } {\text{Tr}\left( {\mathop {\bf{F}}\nolimits_{\mathop k\nolimits_l^d } {\mathop {\bf{F}}\nolimits_{\mathop k\nolimits_l^d } ^H}} \right)}  \le \mathop P\nolimits_B .
		\end{equation}\label{}\noindent 
		 Similarly, the transmit signal of the ${\mathop k\nolimits_l^u }$th UL user is given by
		\begin{equation}
		\mathop {\bf{x}}\nolimits_{\mathop k\nolimits_l^u }  = \mathop {\bf{F}}\nolimits_{\mathop k\nolimits_l^u } \mathop {\bf{s}}\nolimits_{\mathop k\nolimits_l^u } ,  
		\end{equation}\label{}\noindent 
		where $\mathop {\mathbf{s}}\nolimits_{\mathop k\nolimits_l^u }  \in {\mathbb{C}^{\mathop b\nolimits_u  \times 1}} $ denotes the data symbol vector with unit power, transmitted by the ${\mathop k\nolimits_l^u }$th UL user to the  $l{\text{th}}$ BS, and $\mathop {\mathbf{F}}\nolimits_{\mathop k\nolimits_l^u } $ is the related TPC matrix, meeting the following power constraint
		\begin{equation}
		{\text{Tr}}\left( {\mathop {\mathbf{F}}\nolimits_{\mathop k\nolimits_l^u } {\mathop {\mathbf{F}}\nolimits_{\mathop k\nolimits_l^u } ^H}} \right) \leq \mathop P\nolimits_U .
		\end{equation}\label{}	
		\indent Besides, we define the configuration of the $m$th reflecting element on RIS as ${\varphi _m} = {e^{j{\theta _m}}}$, where $\theta_m\in\left[0,2\pi\right)$ is the corresponding phase shift, and $m \in \left\{ {1, \cdots ,M} \right\} \buildrel \Delta \over = \mathcal{M}$. Hence, the configuration vector of RIS is 
		$\bm{\phi}  = {\left[ {{\varphi _1},\cdots,{\varphi _m},\cdots,{\varphi _M}} \right]^T}$, which can be expressed equivalently in a matrix form as ${\bf{\Phi }} = {\rm{diag}}\left( \bm{\phi}  \right)$. For convenience, all end-to-end wireless channels are denoted as  ${{\bf{H}}_{recei,trans}}$. The first subscript represents the receiver, while the second one indicates the transmitter. Particularly, the baseband channels from the  $j{\text{th}}$ BS to the ${\mathop k\nolimits_l^d }$th DL user  and the $l{\text{th}}$ BS are denoted by  $\mathop {\mathbf{H}}\nolimits_{\mathop k\nolimits_l^d ,j} {\text{ }} \in {\mathbb{C}^{\mathop M\nolimits_U^r  \times \mathop M\nolimits_B^t }}$ and $\mathop {\mathbf{H}}\nolimits_{l,j}  \in {\mathbb{C}^{\mathop M\nolimits_B^r  \times \mathop M\nolimits_B^t }}$, respectively. The channels from the ${\mathop k\nolimits_l^u }$th UL user to the $j{\text{th}}$ BS and ${\mathop i\nolimits_j^d }{\text{th}}$ DL user are denoted by $\mathop {\mathbf{H}}\nolimits_{j,\mathop k\nolimits_l^u }  \in {\mathbb{C}^{\mathop M\nolimits_B^r  \times \mathop M\nolimits_U^t }}$ and $\mathop {\mathbf{H}}\nolimits_{\mathop i\nolimits_j^d ,\mathop k\nolimits_l^u } {\text{ }} \in {\mathbb{C}^{\mathop M\nolimits_U^r  \times \mathop M\nolimits_U^t }} $, respectively. The channels from the RIS to the $l{\text{th}}$ BS and ${\mathop k\nolimits_l^d }$th DL user, from the $l{\text{th}}$ BS and ${\mathop k\nolimits_l^u }$th UL user to the RIS are denoted by $\mathop {\mathbf{G}}\nolimits_{l,R}  \in {\mathbb{C}^{\mathop M\nolimits_B^r  \times M}}$, 
		 $\mathop {\mathbf{G}}\nolimits_{\mathop k\nolimits_l^d ,R}  \in {\mathbb{C}^{\mathop M\nolimits_U^r  \times M}}$, $\mathop {\mathbf{G}}\nolimits_{R,l}  \in {\mathbb{C}^{M \times \mathop M\nolimits_B^t }}$, and  $\mathop {\mathbf{G}}\nolimits_{R,\mathop k\nolimits_l^u {\text{ }}}  \in {\mathbb{C}^{M \times \mathop M\nolimits_U^t }}$, respectively.  {{We assume that the CSI of all channels has been obtained at BSs.\footnote{Though it is challenging to acquire the perfect CSI knowledge considering the abundant reflecting elements, existing works have proposed some practical low-complexity schemes to provide qualified CSI estimation with applicable pilot overheads (see \cite{CascadedChannel} and \cite{DeepLearning} for more details). Besides, our proposed scheme with full CSI can provide a performance upper bound for realistic scenarios and robust designs.}  These BSs calculate the optimal phase matrix cooperatively and send it to the controller of RIS.}}\\		
        Explicitly, the signal received at the $l{\text{th}}$ BS is given by
		\begin{equation}\label{eq:yl}
			\begin{split}
	\mathop {\bf{y}}\nolimits_l  &= \underbrace {\left( {\mathop {\bf{H}}\nolimits_{l,\mathop k\nolimits_l^u }  + \mathop {\bf{G}}\nolimits_{l,R} {\bf{\Phi }}\mathop {\bf{G}}\nolimits_{R,\mathop k\nolimits_l^u } } \right)\mathop {\bf{x}}\nolimits_{\mathop k\nolimits_l^u } }_{{\rm{desired \text{  } signal}}}\\&
	{\rm{ + }}\underbrace {\sum\limits_{\scriptstyle i = 1\hfill\atop
			\scriptstyle i \ne k\hfill}^{\mathop K\nolimits_l^u } {\left( {\mathop {\bf{H}}\nolimits_{l,\mathop i\nolimits_l^u }  + \mathop {\bf{G}}\nolimits_{l,R} {\bf{\Phi }}\mathop {\bf{G}}\nolimits_{R,\mathop i\nolimits_l^u } } \right)\mathop {\bf{x}}\nolimits_{\mathop i\nolimits_l^u } } }_{{\rm{intra-cell~ interference~from~uplink}}}\\&
		{\rm{ + }}\underbrace {\sum\limits_{\scriptstyle j = 1\hfill\atop
			\scriptstyle j \ne l\hfill}^L {\sum\limits_{i = 1}^{\mathop K\nolimits_j^u } {\left( {\mathop {\bf{H}}\nolimits_{l,\mathop i\nolimits_j^u }  + \mathop {\bf{G}}\nolimits_{l,R} {\bf{\Phi }}\mathop {\bf{G}}\nolimits_{R,\mathop i\nolimits_j^u } } \right)\mathop {\bf{x}}\nolimits_{\mathop i\nolimits_j^u } } } }_{{\rm{inter-cell~interference~from~neighbor~UL~users}}}\\&
		+ \underbrace {\sqrt {\mathop \rho \nolimits_{l,l}^{ - 1} } \left( {\mathop {\bf{H}}\nolimits_{l,l}  + \mathop {\bf{G}}\nolimits_{l,R} {\bf{\Phi }}\mathop {\bf{G}}\nolimits_{R,l} } \right)\sum\limits_{i = 1}^{\mathop K\nolimits_l^d } {\mathop {\bf{x}}\nolimits_{\mathop i\nolimits_l^d } } }_{{\rm{self - interference}}} \\&		
		{\rm{ + }}\underbrace {\sum\limits_{\scriptstyle j = 1\hfill\atop
				\scriptstyle j \ne l\hfill}^L {\left( {\mathop {\bf{H}}\nolimits_{l,j}  + \mathop {\bf{G}}\nolimits_{l,R} {\bf{\Phi }}\mathop {\bf{G}}\nolimits_{R,j} } \right)} \sum\limits_{i = 1}^{\mathop K\nolimits_j^d } {\mathop {\bf{x}}\nolimits_{\mathop i\nolimits_j^d } } }_{{\rm{inter-cell~interference~from~neighbor~BSs}}}
		{\rm{ + }}\mathop {\bf{n}}\nolimits_l {\rm{ }},
\end{split}
		\end{equation}
		\noindent where $ \rho_{l,l}> {\text{1}}  $ is the SIC coefficient, and ${{\bf{n}}_{l}}$ is the AWGN vector following the distribution of ${\cal C}{\cal N}\left( {0,\sigma _{B}^{\rm{2}}\mathop {\bf{I}}\nolimits_{\mathop M\nolimits_B^r } } \right)$. 	Similarly, the signal received at the ${\mathop k\nolimits_l^d }$th DL user is given by
		\begin{equation}\label{eq:ydkl}
			\begin{split}
	\mathop {\bf{y}}\nolimits_{\mathop k\nolimits_l^d }  =& \underbrace {\left( {\mathop {\bf{H}}\nolimits_{\mathop k\nolimits_l^d ,l}  + \mathop {\bf{G}}\nolimits_{\mathop k\nolimits_l^d ,R} {\bf{\Phi }}\mathop {\bf{G}}\nolimits_{R,l} } \right)\mathop {\bf{x}}\nolimits_{\mathop k\nolimits_l^d } }_{{\rm{desired \text{  } signal}}} \\& + \underbrace {\left( {\mathop {\bf{H}}\nolimits_{\mathop k\nolimits_l^d ,l}  + \mathop {\bf{G}}\nolimits_{\mathop k\nolimits_l^d ,R} {\bf{\Phi }}\mathop {\bf{G}}\nolimits_{R,l} } \right)\sum\limits_{\scriptstyle i = 1\hfill\atop
			\scriptstyle i \ne k\hfill}^{\mathop K\nolimits_l^d } {\mathop {\bf{x}}\nolimits_{\mathop i\nolimits_l^d } } }_{{\rm{intra-cell~interference~from~downlink}}} \\& + \underbrace {\sum\limits_{\scriptstyle j = 1\hfill\atop
			\scriptstyle j \ne l\hfill}^L {\left( {\mathop {\bf{H}}\nolimits_{\mathop k\nolimits_l^d ,j}  + \mathop {\bf{G}}\nolimits_{\mathop k\nolimits_l^d ,R} {\bf{\Phi }}\mathop {\bf{G}}\nolimits_{R,j} } \right)} \sum\limits_{i = 1}^{\mathop K\nolimits_j^d } {\mathop {\bf{x}}\nolimits_{\mathop i\nolimits_j^d } } }_{{\rm{inter-cell~interference~from~neighbor~BSs}}} \\&
		+ \underbrace {\sum\limits_{i = 1}^{\mathop K\nolimits_l^u } {\left( {\mathop {\bf{H}}\nolimits_{\mathop k\nolimits_l^d ,\mathop i\nolimits_l^u }  + \mathop {\bf{G}}\nolimits_{\mathop k\nolimits_l^d ,R} {\bf{\Phi }}\mathop {\bf{G}}\nolimits_{R,\mathop i\nolimits_l^u } } \right)\mathop {\bf{x}}\nolimits_{\mathop i\nolimits_l^u } } }_{{\rm{intra-cell~interference~from~uplink}}}
		 \\&+ \underbrace {\sum\limits_{\scriptstyle j = 1\hfill\atop
		 		\scriptstyle j \ne l\hfill}^L {\sum\limits_{i = 1}^{\mathop K\nolimits_j^u } {\left( {\mathop {\bf{H}}\nolimits_{\mathop k\nolimits_l^d ,\mathop i\nolimits_j^u }  + \mathop {\bf{G}}\nolimits_{\mathop k\nolimits_l^d ,R} {\bf{\Phi }}\mathop {\bf{G}}\nolimits_{R,\mathop i\nolimits_j^u } } \right)\mathop {\bf{x}}\nolimits_{\mathop i\nolimits_j^u } } } }_{{\rm{inter-cell~interference~from~neighbor~UL~users}}}  + \mathop {\bf{n}}\nolimits_{\mathop k\nolimits_l^d } ,
			\end{split}
		\end{equation}
		where $\mathop {\bf{n}}\nolimits_{\mathop k\nolimits_l^d } $ denotes the AWGN at the user side and obeys the distribution of $\mathcal{C}\mathcal{N} ( {0,\mathop \sigma \nolimits_U^2 \mathop {\bf{I}}\nolimits_{\mathop M\nolimits_U^r } } )$. { From (\ref{eq:yl}) and (\ref{eq:ydkl}) we can find that, the interference environment confronted is complicated, which poses a great challenge to improve the system performance.}  To simply the expressions, let us define $\mathop {{\bf{\bar H}}}\nolimits_{\mathop k\nolimits_l^d ,j} {\rm{ = }}\mathop {\bf{H}}\nolimits_{\mathop k\nolimits_l^d ,j}  + \mathop {\bf{G}}\nolimits_{\mathop k\nolimits_l^d ,R} {\bf{\Phi }}\mathop {\bf{G}}\nolimits_{R,j} $,  $\mathop {{\bf{\bar H}}}\nolimits_{l,j} {\rm{ = }}\mathop {\bf{H}}\nolimits_{l,j}  + \mathop {\bf{G}}\nolimits_{l,R} {\bf{\Phi }}\mathop {\bf{G}}\nolimits_{R,j} $,  $\mathop {{\bf{\bar H}}}\nolimits_{l,\mathop i\nolimits_j^u } {\rm{ = }}\mathop {\bf{H}}\nolimits_{l,\mathop i\nolimits_j^u }  + \mathop {\bf{G}}\nolimits_{l,R} {\bf{\Phi }}\mathop {\bf{G}}\nolimits_{R,\mathop i\nolimits_j^u }  $  and $\mathop {{\bf{\bar H}}}\nolimits_{\mathop k\nolimits_l^d ,\mathop i\nolimits_j^u } {\rm{ = }}\mathop {\bf{H}}\nolimits_{\mathop k\nolimits_l^d ,\mathop i\nolimits_j^u }  + \mathop {\bf{G}}\nolimits_{\mathop k\nolimits_l^d ,R} {\bf{\Phi }}\mathop {\bf{G}}\nolimits_{R,\mathop i\nolimits_j^u } $  as the equivalent channels correspondingly, which are the superposition of the direct and reflected links. Consequently, ${{\mathbf{y}}_{{l}}}$ can be rewritten briefly as 
		\begin{equation}\label{}
			\begin{split}
\mathop {\bf{y}}\nolimits_l  =& \sum\limits_{j = 1}^L {\sqrt {\mathop \rho \nolimits_{l,j}^{ - 1} } \mathop {{\bf{\bar H}}}\nolimits_{l,j} } \sum\limits_{i = 1}^{\mathop K\nolimits_j^d } {\mathop {\bf{F}}\nolimits_{\mathop i\nolimits_j^d } \mathop {\bf{s}}\nolimits_{\mathop i\nolimits_j^d }   }  + \sum\limits_{j = 1}^L {\sum\limits_{i = 1}^{\mathop K\nolimits_j^u } {\mathop {{\bf{\bar H}}}\nolimits_{l,\mathop i\nolimits_j^u } \mathop {\bf{F}}\nolimits_{\mathop i\nolimits_j^u } \mathop {\bf{s}}\nolimits_{\mathop i\nolimits_j^u } } }  + \mathop {\bf{n}}\nolimits_l ,
			\end{split}
		\end{equation}
where coefficient ${\rho _{l,j}}$ is defined as
		\begin{equation*}\label{}
{{\rho _{{l,j}}} = \left\{ \begin{gathered}
	{\rho _{{l,l}}},\quad {\text{if}}{\text{ }}j = l; \hfill \\
	1,\quad{\text{ }}{\text{ }}{\text{    otherwise}}{\text{. }} \hfill \\ 
\end{gathered}  \right.}
\end{equation*}
Similarly, $\mathop {\bf{y}}\nolimits_{\mathop k\nolimits_l^d } $ can be reformulated as
		\begin{equation}\label{}
			\begin{split}
\mathop {\bf{y}}\nolimits_{\mathop k\nolimits_l^d }  = \sum\limits_{j = 1}^L {\mathop {{\bf{\bar H}}}\nolimits_{\mathop k\nolimits_l^d ,j} } \sum\limits_{i = 1}^{\mathop K\nolimits_j^d } {\mathop {\bf{F}}\nolimits_{\mathop i\nolimits_j^d } \mathop {\bf{s}}\nolimits_{\mathop i\nolimits_j^d } }  + \sum\limits_{j = 1}^L {\sum\limits_{i = 1}^{\mathop K\nolimits_j^u } {\mathop {{\bf{\bar H}}}\nolimits_{\mathop k\nolimits_l^d ,\mathop i\nolimits_j^u } \mathop {\bf{F}}\nolimits_{\mathop i\nolimits_j^u } \mathop {\bf{s}}\nolimits_{\mathop i\nolimits_j^u } } }  + \mathop {\bf{n}}\nolimits_{\mathop k\nolimits_l^d } {\rm{   }}.
			\end{split}
		\end{equation}
Then, the achievable data rate (bit/s/Hz) at the ${\mathop k\nolimits_l^u }$th UL user is
		\begin{equation}\label{Eqa:Rukl}
\begin{gathered}
\mathop R\nolimits_{\mathop k\nolimits_l^u } \left( {{\bf{F,\Phi }}} \right) = \log \det \left( {\mathop {\bf{I}}\nolimits_{\mathop M\nolimits_B^r }  + \mathop {{\bf{\bar H}}}\nolimits_{l,\mathop k\nolimits_l^u } \mathop {\bf{F}}\nolimits_{\mathop k\nolimits_l^u } \mathop {\bf{F}}\nolimits_{\mathop k\nolimits_l^u }^H {\mathop {{\bf{\bar H}}}\nolimits_{l,\mathop k\nolimits_l^u } ^H}\mathop {\bf{V}}\nolimits_{\mathop k\nolimits_l^u }^{ - 1} } \right),
\end{gathered} 
		\end{equation}
where ${\bf{F}} = \left\{ {\mathop {\bf{F}}\nolimits_{\mathop k\nolimits_l^d } ,\mathop {\bf{F}}\nolimits_{\mathop k\nolimits_l^u } ;\forall k,l} \right\}$ includes all UL and DL TPC matrices, and 
\begin{equation}\label{}
				\begin{split}
\mathop {\bf{V}}\nolimits_{\mathop k\nolimits_l^u }  = & \mathop {\sum\limits_{j = 1}^L {} \sum\limits_{i = 1}^{\mathop K\nolimits_j^u } {} }\limits_{\left( {i,j \ne k,l} \right)} \mathop {{\bf{\bar H}}}\nolimits_{l,\mathop i\nolimits_j^u } \mathop {\bf{F}}\nolimits_{\mathop i\nolimits_j^u } {\mathop {\bf{F}}\nolimits_{\mathop i\nolimits_j^u } ^H}{\mathop {{\bf{\bar H}}}\nolimits_{l,\mathop i\nolimits_j^u } ^H} \\& + \sum\limits_{j = 1}^L {\sum\limits_{i = 1}^{\mathop K\nolimits_j^d } {\mathop \rho \nolimits_{l,j}^{ - 1} } \mathop {{\bf{\bar H}}}\nolimits_{l,j} \mathop {\bf{F}}\nolimits_{\mathop i\nolimits_j^d } {{\mathop {\bf{F}}\nolimits_{\mathop i\nolimits_j^d } }^H}{\mathop {{\bf{\bar H}}}\nolimits_{l,j} ^H}}  + \mathop \sigma \nolimits_B^2 \mathop {\bf{I}}\nolimits_{\mathop M\nolimits_B^r } ,
			\end{split}
		\end{equation}
represents the interference-plus-noise covariance matrix at the ${\mathop k\nolimits_l^u }$th UL user. Similarly, the achievable data rate at the ${\mathop k\nolimits_l^d }$th  DL user is given by
		\begin{equation}\label{Eqa:Rdkl}
\begin{gathered}
\mathop R\nolimits_{\mathop k\nolimits_l^d } \left( {{\bf{F,\Phi }}} \right) = \log \det \left( {\mathop {\bf{I}}\nolimits_{\mathop M\nolimits_U^r }  + \mathop {{\bf{\bar H}}}\nolimits_{\mathop k\nolimits_l^d ,l} \mathop {\bf{F}}\nolimits_{\mathop k\nolimits_l^d } \mathop {\bf{F}}\nolimits_{\mathop k\nolimits_l^d }^H \mathop {{\bf{\bar H}}}\nolimits_{\mathop k\nolimits_l^d ,l}^H \mathop {\bf{V}}\nolimits_{\mathop k\nolimits_l^d }^{ - 1} } \right) ,
\end{gathered} 
		\end{equation}
\noindent where $\mathop {\bf{V}}\nolimits_{\mathop k\nolimits_l^d } $ is the interference-plus-noise covariance matrix at the ${\mathop k\nolimits_l^d }$th DL user and is given by
		\begin{equation}\label{}
			\begin{split}			
\mathop {\bf{V}}\nolimits_{\mathop k\nolimits_l^d }  =& \mathop {\sum\limits_{j = 1}^L {} \sum\limits_{i = 1}^{\mathop K\nolimits_j^d } {} }\limits_{\left( {i,j \ne k,l} \right)} \mathop {{\bf{\bar H}}}\nolimits_{\mathop k\nolimits_l^d ,j} \mathop {\bf{F}}\nolimits_{\mathop i\nolimits_j^d } \mathop {\bf{F}}\nolimits_{\mathop i\nolimits_j^d }^H \mathop {{\bf{\bar H}}}\nolimits_{\mathop k\nolimits_l^d ,j}^H  \\& + \sum\limits_{j = 1}^L {\sum\limits_{i = 1}^{\mathop K\nolimits_j^u } {\mathop {{\bf{\bar H}}}\nolimits_{\mathop k\nolimits_l^d ,\mathop i\nolimits_j^u } \mathop {\bf{F}}\nolimits_{\mathop i\nolimits_j^u } } {{\mathop {\bf{F}}\nolimits_{\mathop i\nolimits_j^u } }^H}{\mathop {{\bf{\bar H}}}\nolimits_{\mathop k\nolimits_l^d ,\mathop i\nolimits_j^u } ^H}}  + \mathop \sigma \nolimits_U^2 \mathop {\bf{I}}\nolimits_{\mathop M\nolimits_U^r } .
			\end{split}	
		\end{equation}
		\subsection{Problem Formulation}
		To achieve the full potential of FD mobile networking, in this paper, we aim to maximize the SR of all users and BSs by jointly optimizing the overall TPC matrices ${\bf{F}}$  and the phase shift configuration ${\bf{\Phi }}$ at RIS, subject to their corresponding power budget and unit modulus constraint. Explicitly, the overall problem is formulated as follows
		\begin{subequations}\label{OF:original}
			\begin{align}\label{OF:originala}
					& \mathop {\max }\limits_{{\bf{F,\Phi }}} {\rm{     }}\sum\limits_{l = 1}^L {\sum\limits_{k = 1}^{\mathop K\nolimits_l^d } {\mathop R\nolimits_{\mathop k\nolimits_l^d } } \left( {{\bf{F,\Phi }}} \right)} {\rm{ + }}\sum\limits_{l = 1}^L {\sum\limits_{k = 1}^{\mathop K\nolimits_l^u } {\mathop R\nolimits_{\mathop k\nolimits_l^u } } \left( {{\bf{F,\Phi }}} \right)} {\rm{     }}\\
					&s.t.\quad {\rm{     Tr}}\left( {\mathop {\bf{F}}\nolimits_{\mathop k\nolimits_l^u } {\mathop {\bf{F}}\nolimits_{\mathop k\nolimits_l^u } ^H}} \right) \le \mathop P\nolimits_U {\rm{,  }}\forall k,l,\\
					&\qquad{\rm{          }}\sum\limits_{k = 1}^{\mathop K\nolimits_l^d } {{\rm{     Tr}}\left( {\mathop {\bf{F}}\nolimits_{\mathop k\nolimits_l^d } {\mathop {\bf{F}}\nolimits_{\mathop k\nolimits_l^d } ^H}} \right)}  \le \mathop P\nolimits_B ,{\rm{  }}\forall l,\\
					&\qquad{\rm{          }}0 \le \theta  \le 2\pi ,\left| {\mathop \varphi \nolimits_m } \right| = 1,\forall m .
			\end{align}
		\end{subequations}
		It can be observed that above problem is hard be solve for the reason that the overall TPC matrices $\bf F$  and the phase shift matrix ${\bf{\Phi }}$ are deeply coupled, as shown in (\ref{Eqa:Rukl}) and (\ref{Eqa:Rdkl}). Additionally, the unit modulus constraint of the phase shift matrix makes the feasible set of (\ref{OF:original}) non-convex. In the following, we resort to a BCD-based method to address this problem.\\
		\indent We make some reformulations with respect to the original problem at first. Inspired by \cite{WeightedMMSEApproach}, we exploit the relationship between the SR and MMSE to transform (\ref{OF:original}) into a more tractable one. Particularly, upon introducing the decoding matrices, the estimated signals of uplink and downlink are respectively given by
		\begin{equation}\label{}
			\begin{split}
\mathop {{\bf{\hat s}}}\nolimits_{\mathop k\nolimits_l^u }  = \mathop {\bf{U}}\nolimits_{\mathop k\nolimits_l^u } \mathop {\bf{y}}\nolimits_l {\rm{    ,     }}\mathop {{\bf{\hat s}}}\nolimits_{\mathop k\nolimits_l^d }  = \mathop {\bf{U}}\nolimits_{\mathop k\nolimits_l^d } \mathop {\bf{y}}\nolimits_{\mathop k\nolimits_l^d } {\rm{ }}.
			\end{split}
		\end{equation}
 where $\mathop {\mathbf{U}}\nolimits_{\mathop k\nolimits_l^u }  \in {\mathbb{C}^{\mathop b\nolimits_u  \times \mathop M\nolimits_B^r }}$  and $\mathop {\mathbf{U}}\nolimits_{\mathop k\nolimits_l^d }  \in {\mathbb{C}^{\mathop b\nolimits_d  \times \mathop M\nolimits_U^r }}$ are the  decoding matrices {for} the UL and DL users, respectively. Therefore, the mean square error (MSE) of the estimated signals are expressed as		
	\begin{equation}\label{Eqa:eB}
	\begin{split}
		\mathop {\mathbf{E}}\nolimits_{\mathop k\nolimits_l^u } = & \mathbb{E}\left[ {\left( {\mathop {{\mathbf{\hat s}}}\nolimits_{\mathop k\nolimits_l^u }  - \mathop {\mathbf{s}}\nolimits_{\mathop k\nolimits_l^u } } \right){{\left( {\mathop {{\mathbf{\hat s}}}\nolimits_{\mathop k\nolimits_l^u }  - \mathop {\mathbf{s}}\nolimits_{\mathop k\nolimits_l^u } } \right)}^H}} \right] \hfill \\
	=	& \left( {\mathop {\mathbf{U}}\nolimits_{\mathop k\nolimits_l^u } \mathop {{\mathbf{\bar H}}}\nolimits_{l,\mathop k\nolimits_l^u } \mathop {\mathbf{F}}\nolimits_{\mathop k\nolimits_l^u }  - {{\mathbf{I}}_{\mathop b\nolimits_u }}} \right){\left( {\mathop {\mathbf{U}}\nolimits_{\mathop k\nolimits_l^u } \mathop {{\mathbf{\bar H}}}\nolimits_{l,\mathop k\nolimits_l^u } \mathop {\mathbf{F}}\nolimits_{\mathop k\nolimits_l^u }  - {{\mathbf{I}}_{\mathop b\nolimits_u }}} \right)^H} \\&+ \mathop {\mathbf{U}}\nolimits_{\mathop k\nolimits_l^u } \mathop {\mathbf{V}}\nolimits_{\mathop k\nolimits_l^u } {\mathop {\mathbf{U}}\nolimits_{\mathop k\nolimits_l^u } ^H} \hfill ,
	\end{split}	
\end{equation}	
		\begin{equation}\label{Eqa:eU}
	\begin{split}			
	\mathop {\mathbf{E}}\nolimits_{\mathop k\nolimits_l^d }  =& \mathbb{E}\left[ {\left( {\mathop {{\mathbf{\hat s}}}\nolimits_{\mathop k\nolimits_l^d }  - \mathop {\mathbf{s}}\nolimits_{\mathop k\nolimits_l^d } } \right){{\left( {\mathop {{\mathbf{\hat s}}}\nolimits_{\mathop k\nolimits_l^d }  - \mathop {\mathbf{s}}\nolimits_{\mathop k\nolimits_l^d } } \right)}^H}} \right] \hfill \\
	=& \left( {\mathop {\mathbf{U}}\nolimits_{\mathop k\nolimits_l^d } \mathop {{\mathbf{\bar H}}}\nolimits_{\mathop k\nolimits_l^d ,l} \mathop {\mathbf{F}}\nolimits_{\mathop k\nolimits_l^d }  - {{\mathbf{I}}_{\mathop b\nolimits_d }}} \right){\left( {\mathop {\mathbf{U}}\nolimits_{\mathop k\nolimits_l^d } \mathop {{\mathbf{\bar H}}}\nolimits_{\mathop k\nolimits_l^d ,l} \mathop {\mathbf{F}}\nolimits_{\mathop k\nolimits_l^d }  - {{\mathbf{I}}_{\mathop b\nolimits_d }}} \right)^H} \\&+ \mathop {\mathbf{U}}\nolimits_{\mathop k\nolimits_l^d } \mathop {\mathbf{V}}\nolimits_{\mathop k\nolimits_l^d } {\mathop {\mathbf{U}}\nolimits_{\mathop k\nolimits_l^d } ^H} \hfill  .
\end{split}	
\end{equation}		
We further introduce the weighting matrices
  $\mathop {\mathbf{W}}\nolimits_{\mathop k\nolimits_l^u }  \in {\mathbb{C}^{\mathop b\nolimits_u  \times \mathop b\nolimits_u }}$ and $\mathop {\mathbf{W}}\nolimits_{\mathop k\nolimits_l^d }  \in {\mathbb{C}^{\mathop b\nolimits_d  \times \mathop b\nolimits_d }}$  for the UL and DL users, respectively. 
Then (\ref{Eqa:Rukl}) and (\ref{Eqa:Rdkl}) are transformed into the following forms as
		\begin{equation}\label{Eqa:rB}
			\begin{split}
\mathop r\nolimits_{\mathop k\nolimits_l^u } \left( {{\mathbf{F}},{\mathbf{\Phi }},\mathop {\mathbf{U}}\nolimits_{\mathop k\nolimits_l^u } ,\mathop {\mathbf{W}}\nolimits_{\mathop k\nolimits_l^u } } \right) =& \log \det \left( {\mathop {\mathbf{W}}\nolimits_{\mathop k\nolimits_l^u } } \right) - {\text{Tr}}\left( {\mathop {\mathbf{W}}\nolimits_{\mathop k\nolimits_l^u } \mathop {\mathbf{E}}\nolimits_{\mathop k\nolimits_l^u } } \right) \\& + \mathop b\nolimits_u ,
			\end{split}
		\end{equation}
		\begin{equation}\label{Eqa:rU}
			\begin{split}
\mathop r\nolimits_{\mathop k\nolimits_l^d } \left( {{\mathbf{F}},{\mathbf{\Phi }},\mathop {\mathbf{U}}\nolimits_{\mathop k\nolimits_l^d } ,\mathop {\mathbf{W}}\nolimits_{\mathop k\nolimits_l^d } } \right) =& \log \det \left( {\mathop {\mathbf{W}}\nolimits_{\mathop k\nolimits_l^d } } \right) - {\text{Tr}}\left( {\mathop {\mathbf{W}}\nolimits_{\mathop k\nolimits_l^d } \mathop {\mathbf{E}}\nolimits_{\mathop k\nolimits_l^d } } \right) \\&+ \mathop b\nolimits_d {\text{  }}.
			\end{split}
		\end{equation}\\
Therefore, the original problem (13) can be transformed into a more tractable one as follows
\begin{subequations}\label{OF:MMSE}
		\begin{align}\label{OF:MMSEa}
	&\mathop {\max }\limits_{{\mathbf{F,\Phi }},{\bf{U}},{\bf{W}}} {\text{     }}\sum\limits_{l = 1}^L {\sum\limits_{k = 1}^{\mathop K\nolimits_l^d } {\mathop r\nolimits_{\mathop k\nolimits_l^d } } } {\text{ + }}\sum\limits_{l = 1}^L {\sum\limits_{k = 1}^{\mathop K\nolimits_l^u } {\mathop r\nolimits_{\mathop k\nolimits_l^u } } } {\text{     }} \hfill \tag{19a} \\
	&s.t. \qquad \quad {\text{     Tr}}\left( {\mathop {\mathbf{F}}\nolimits_{\mathop k\nolimits_l^u } {\mathop {\mathbf{F}}\nolimits_{\mathop k\nolimits_l^u } ^H}} \right) \leq \mathop P\nolimits_U {\text{  }}\forall k,l, \hfill  \tag{19b}\\
	&\qquad \quad \quad {\text{          }}\sum\limits_{k = 1}^{\mathop K\nolimits_l^d } {\text{Tr}\left( {\mathop {\mathbf{F}}\nolimits_{\mathop k\nolimits_l^d } {\mathop {\mathbf{F}}\nolimits_{\mathop k\nolimits_l^d } ^H}} \right)}  \leq \mathop P\nolimits_B ,{\text{  }}\forall l, \hfill \tag{19c} \\
	&\qquad \quad \quad {\text{          }}0 \leq \theta  \leq 2\pi ,\left| {\mathop \varphi \nolimits_m } \right| = 1,{\text{   }}\forall m .\hfill \tag{19d} 
	\end{align}
\end{subequations}
where ${\bf{U}} = \left\{ {\mathop {\bf{U}}\nolimits_{\mathop k\nolimits_l^d } ,\mathop {\bf{U}}\nolimits_{\mathop k\nolimits_l^u } ;\forall k,l} \right\}$ and ${\bf{W}} = \left\{ {\mathop {\bf{W}}\nolimits_{\mathop k\nolimits_l^d } ,\mathop {\bf{W}}\nolimits_{\mathop k\nolimits_l^u } ;\forall k,l} \right\}$. {We notice that the problem (\ref{OF:MMSE}) is equivalent to the original problem (\ref{OF:original}) since they have the identical optimal solutions to $\mathbf{F}$ and $\mathbf{\Phi}$ \cite[Theorem 1]{WeightedMMSEApproach}. Furthermore, (\ref{OF:MMSEa}) is a convex function for each group of the variables. When all variables except ${\mathop {\mathbf{U}}\nolimits_{\mathop k\nolimits_l^u } }$ are fixed, we can obtain the optimal solution of it directly. Similarly, we obtain the optimal ${\mathop {\mathbf{U}}\nolimits_{\mathop k\nolimits_l^d } }$, ${\mathop {\mathbf{W}}\nolimits_{\mathop k\nolimits_l^u } }$ and $\mathop {\mathbf{W}}\nolimits_{\mathop k\nolimits_l^d } $. Their solutions are shown as below}
			\begin{equation}\label{Eqa:miuB}
	\begin{split}
		\mathop {\mathbf{U}}\nolimits_{\mathop k\nolimits_l^u }  = {\mathop {\mathbf{F}}\nolimits_{\mathop k\nolimits_l^u } ^H}{\mathop {{\mathbf{\bar H}}}\nolimits_{l,\mathop k\nolimits_l^u } ^H}{\left[ {\mathop {{\mathbf{\bar H}}}\nolimits_{l,\mathop k\nolimits_l^u } \mathop {\mathbf{F}}\nolimits_{\mathop k\nolimits_l^u } {\mathop {\mathbf{F}}\nolimits_{\mathop k\nolimits_l^u } ^H}{\mathop {{\mathbf{\bar H}}}\nolimits_{l,\mathop k\nolimits_l^u } ^H} + \mathop {\mathbf{V}}\nolimits_{\mathop k\nolimits_l^u } } \right]^{ - 1}},
	\end{split}
\end{equation}
	\begin{equation}\label{Eqa:miuU}
		\begin{split}
\mathop {\mathbf{U}}\nolimits_{\mathop k\nolimits_l^d }  = {\mathop {\mathbf{F}}\nolimits_{\mathop k\nolimits_l^d } ^H}{\mathop {{\mathbf{\bar H}}}\nolimits_{\mathop k\nolimits_l^d ,l} ^H}{\left[ {\mathop {{\mathbf{\bar H}}}\nolimits_{\mathop k\nolimits_l^d ,l} \mathop {\mathbf{F}}\nolimits_{\mathop k\nolimits_l^d } {\mathop {\mathbf{F}}\nolimits_{\mathop k\nolimits_l^d } ^H}{\mathop {{\mathbf{\bar H}}}\nolimits_{\mathop k\nolimits_l^d ,l} ^H} + \mathop {\mathbf{V}}\nolimits_{\mathop k\nolimits_l^d } } \right]^{ - 1}}{\text{ }},
		\end{split}
	\end{equation}
\begin{equation}\label{Eqa:wB,wU}
\mathop {\mathbf{W}}\nolimits_{\mathop k\nolimits_l^d }  = {\mathop {\mathbf{E}}\nolimits_{\mathop k\nolimits_l^d } ^{ - 1}}{\text{,     }}\mathop {\mathbf{W}}\nolimits_{\mathop k\nolimits_l^u }  = {\mathop {\mathbf{E}}\nolimits_{\mathop k\nolimits_l^u } ^{ - 1}}.
		\end{equation}
	
 In the following, we focus on solving (\ref{OF:MMSE}) with respect to $\mathbf{F}$ and $\mathbf{\Phi}$ alternately, where other variables are obtained according to \eqref{Eqa:miuB}-\eqref{Eqa:wB,wU}. 
\section{TPC Matrices Optimization}
		In this section, we aim to optimize the overall TPC matrices ${\bf{F}}$, while fixing the other variables in \eqref{OF:MMSE}. {Specifically, by substituting (\ref{Eqa:eB}) and (\ref{Eqa:eU}) into the OF of (\ref{OF:MMSE}) and removing the constant and irrelative terms, we obtain a new OF with respect to $\bf{F}$, where all $\left\{ {\mathop {\bf{F}}\nolimits_{\mathop k\nolimits_l^d } ,\mathop {\bf{F}}\nolimits_{\mathop k\nolimits_l^u } ;\forall k,l} \right\}$ are uncoupled. 
	   Explicitly, the optimization problem with respect to the overall TPC matrices ${\bf{F}}$ is given by }
		
		\begin{small}
\begin{subequations}\label{OF:TPC}
	\begin{align}\label{OF:TPCa}
	&\mathop {{\text{min}}}\limits_{\mathbf{F}} \sum\limits_{l = 1}^L {} \sum\limits_{k = 1}^{\mathop K\nolimits_l^d } {} \left( {{\text{Tr}}\left( {{\mathop {\mathbf{F}}\nolimits_{\mathop k\nolimits_l^d } ^H}\mathop {\mathbf{A}}\nolimits_l \mathop {\mathbf{F}}\nolimits_{\mathop k\nolimits_l^d } } \right) - 2\operatorname{Re} {\text{Tr}}\left( {\mathop {\mathbf{W}}\nolimits_{\mathop k\nolimits_l^d } \mathop {\mathbf{U}}\nolimits_{\mathop k\nolimits_l^d } \mathop {{\mathbf{\bar H}}}\nolimits_{\mathop k\nolimits_l^d ,l} \mathop {\mathbf{F}}\nolimits_{\mathop k\nolimits_l^d } } \right)} \right) \nonumber \\&+ \sum\limits_{l = 1}^L {\sum\limits_{k = 1}^{\mathop K\nolimits_l^u } {} } \left( {{\text{Tr}}({\mathop {\mathbf{F}}\nolimits_{\mathop k\nolimits_l^u } ^H}\mathop {\mathbf{A}}\nolimits_{\mathop k\nolimits_l^u } \mathop {\mathbf{F}}\nolimits_{\mathop k\nolimits_l^u } ) - 2\operatorname{Re} {\text{Tr}}\left( {\mathop {\mathbf{W}}\nolimits_{\mathop k\nolimits_l^u } \mathop {\mathbf{U}}\nolimits_{\mathop k\nolimits_l^u } \mathop {{\mathbf{\bar H}}}\nolimits_{l,\mathop k\nolimits_l^u } \mathop {\mathbf{F}}\nolimits_{\mathop k\nolimits_l^u } } \right)} \right) \hfill  \\
&s.t. \quad {\text{     Tr}}\left( {\mathop {\mathbf{F}}\nolimits_{\mathop k\nolimits_l^u } {\mathop {\mathbf{F}}\nolimits_{\mathop k\nolimits_l^u } ^H}} \right) \leq \mathop P\nolimits_U {\text{  }}\forall k,l, \hfill \\
&\quad \quad {\text{          }}\sum\limits_{k = 1}^{\mathop K\nolimits_l^d } {{\text{Tr}}\left( {\mathop {\mathbf{F}}\nolimits_{\mathop k\nolimits_l^d } {\mathop {\mathbf{F}}\nolimits_{\mathop k\nolimits_l^d } ^H}} \right)}  \leq \mathop P\nolimits_B ,{\text{  }}\forall l. \hfill  
	\end{align}
\end{subequations}
\end{small}

where 
		\begin{equation}\label{Al}
			\begin{split}
\mathop {\bf{A}}\nolimits_l {\rm{ = }}&\sum\limits_{j = 1}^L {\sum\limits_{i = 1}^{\mathop K\nolimits_j^d } {{\mathop {{\bf{\bar H}}}\nolimits_{\mathop i\nolimits_j^d ,l} ^H}{\mathop {\bf{U}}\nolimits_{\mathop i\nolimits_j^d } ^H}\mathop {\bf{W}}\nolimits_{\mathop i\nolimits_j^d } \mathop {\bf{U}}\nolimits_{\mathop i\nolimits_j^d } \mathop {{\bf{\bar H}}}\nolimits_{\mathop i\nolimits_j^d ,l} } }  \\&+ \sum\limits_{j = 1}^L {\sum\limits_{i = 1}^{\mathop K\nolimits_j^u } {\mathop \rho \nolimits_{l,j}^{ - 1} } {\mathop {{\bf{\bar H}}}\nolimits_{j,l} ^H}{\mathop {\bf{U}}\nolimits_{\mathop i\nolimits_j^u } ^H}\mathop {\bf{W}}\nolimits_{\mathop i\nolimits_j^u } \mathop {\bf{U}}\nolimits_{\mathop i\nolimits_j^u } \mathop {{\bf{\bar H}}}\nolimits_{j,l} } ,
			\end{split}
		\end{equation}
			\begin{equation}\label{Aukl}
		\begin{split}
		\mathop {\bf{A}}\nolimits_{\mathop k\nolimits_l^u }  =& \sum\limits_{j = 1}^L {\sum\limits_{i = 1}^{\mathop K\nolimits_j^d } {{\mathop {{\bf{\bar H}}}\nolimits_{\mathop i\nolimits_j^d ,\mathop k\nolimits_l^u } ^H}{\mathop {\bf{U}}\nolimits_{\mathop i\nolimits_j^d } ^H}\mathop {\bf{W}}\nolimits_{\mathop i\nolimits_j^d } \mathop {\bf{U}}\nolimits_{\mathop i\nolimits_j^d } \mathop {{\bf{\bar H}}}\nolimits_{\mathop i\nolimits_j^d ,\mathop k\nolimits_l^u } } }  \\&+ \sum\limits_{j = 1}^L {\sum\limits_{i = 1}^{\mathop K\nolimits_j^u } {{\mathop {{\bf{\bar H}}}\nolimits_{j,\mathop k\nolimits_l^u } ^H}{\mathop {\bf{U}}\nolimits_{\mathop i\nolimits_j^u } ^H}\mathop {\bf{W}}\nolimits_{\mathop i\nolimits_j^u } \mathop {\bf{U}}\nolimits_{\mathop i\nolimits_j^u } \mathop {{\bf{\bar H}}}\nolimits_{j,\mathop k\nolimits_l^u } } }  .
		\end{split}
	\end{equation}
\indent Obviously,  (\ref{OF:TPC}) is a convex problem and can be solved by using the mature CVX toolbox, though the computational complexity may be high especially for a large antenna array in a practical system design. To reduce the complexity, we turn to a Lagrangian multiplier method to obtain a closed-form solution. Specifically, denoting $ \left\{ {\mathop \lambda \nolimits_l  \ge 0,\mathop \lambda \nolimits_{\mathop k\nolimits_l^u }  \ge 0;\forall k,l} \right\}$ as the Lagrangian multipliers associated with the transmitted power constraints, we get the Lagrangian function of (\ref{OF:TPC}) as \par
\vspace{-5mm}
\begin{footnotesize}	
		\begin{equation}\label{OF:larg}
			\begin{split}
	&\mathcal{L} \left( {{\mathbf{F}},\{\lambda_l, \mathop \lambda \nolimits_{\mathop k\nolimits_l^u }\} } \right) =\\& \sum\limits_{l = 1}^L {} \sum\limits_{k = 1}^{\mathop K\nolimits_l^d } {} \left( {{\text{Tr}}\left( {{\mathop {\mathbf{F}}\nolimits_{\mathop k\nolimits_l^d } ^H}\mathop {\mathbf{A}}\nolimits_l \mathop {\mathbf{F}}\nolimits_{\mathop k\nolimits_l^d } } \right) - 2\operatorname{Re} {\text{Tr}}\left( {\mathop {\mathbf{W}}\nolimits_{\mathop k\nolimits_l^d } \mathop {\mathbf{U}}\nolimits_{\mathop k\nolimits_l^d } \mathop {{\mathbf{\bar H}}}\nolimits_{\mathop k\nolimits_l^d ,l} \mathop {\mathbf{F}}\nolimits_{\mathop k\nolimits_l^d } } \right)} \right) \\& + \sum\limits_{l = 1}^L {\sum\limits_{k = 1}^{\mathop K\nolimits_l^u } {} } \left( {{\text{Tr}}({\mathop {\mathbf{F}}\nolimits_{\mathop k\nolimits_l^u } ^H}\mathop {\mathbf{A}}\nolimits_{\mathop k\nolimits_l^u } \mathop {\mathbf{F}}\nolimits_{\mathop k\nolimits_l^u } ) - 2\operatorname{Re} {\text{Tr}}\left( {\mathop {\mathbf{W}}\nolimits_{\mathop k\nolimits_l^u } \mathop {\mathbf{U}}\nolimits_{\mathop k\nolimits_l^u } \mathop {{\mathbf{\bar H}}}\nolimits_{l,\mathop k\nolimits_l^u } \mathop {\mathbf{F}}\nolimits_{\mathop k\nolimits_l^u } } \right)} \right) \hfill \\
	&+ \sum\limits_{l = 1}^L {\sum\limits_{k = 1}^{\mathop K\nolimits_l^u } {} } {\lambda _{\mathop k\nolimits_l^u }}\left( {{\text{Tr}}({\mathop {\mathbf{F}}\nolimits_{\mathop k\nolimits_l^u } ^H}\mathop {\mathbf{F}}\nolimits_{\mathop k\nolimits_l^u } ) - \mathop P\nolimits_U } \right) \\&
	 + \sum\limits_{l = 1}^L {\lambda _l}\left( {\sum\limits_{k = 1}^{\mathop K\nolimits_l^d } {{\text{Tr}}\left( {\mathop {\mathbf{F}}\nolimits_{\mathop k\nolimits_l^d } {\mathop {\mathbf{F}}\nolimits_{\mathop k\nolimits_l^d } ^H}} \right)}  - \mathop P\nolimits_B } \right) \hfill .
			\end{split}
		\end{equation}
	\end{footnotesize} 

\noindent   By taking the derivative of (\ref{OF:larg}) w.r.t. $\left\{ {\mathop {\bf{F}}\nolimits_{\mathop k\nolimits_l^d } ,\mathop {\bf{F}}\nolimits_{\mathop k\nolimits_l^u } }\right\}$ and setting it to be zero, we obtain the optimal solution as a function of Lagrangian multipliers as
		\begin{equation}\label{UpdateFdkl}
\mathop {\mathbf{F}}\nolimits_{\mathop k\nolimits_l^d } \left( {\mathop \lambda \nolimits_l } \right){\text{ = }}{\left( {\mathop {\mathbf{A}}\nolimits_l  + \mathop \lambda \nolimits_l \mathop {\mathbf{I}}\nolimits_{\mathop M\nolimits_B^t } } \right)^{ - 1}}{\mathop {{\mathbf{\bar H}}}\nolimits_{\mathop k\nolimits_l^d ,l} ^H}{\mathop {\mathbf{U}}\nolimits_{\mathop k\nolimits_l^d } ^H}\mathop {\mathbf{W}}\nolimits_{\mathop k\nolimits_l^d } ,
		\end{equation}	
\begin{equation}\label{UpdateFukl}
\mathop {\mathbf{F}}\nolimits_{\mathop k\nolimits_l^u } \left( {\mathop \lambda \nolimits_{\mathop k\nolimits_l^u } } \right) = {\left( {\mathop {\mathbf{A}}\nolimits_{\mathop k\nolimits_l^u }  + \mathop \lambda \nolimits_{\mathop k\nolimits_l^u } \mathop {\mathbf{I}}\nolimits_{\mathop M\nolimits_U^t } } \right)^{ - 1}}{\mathop {{\mathbf{\bar H}}}\nolimits_{l,\mathop k\nolimits_l^u } ^H}{\mathop {\mathbf{U}}\nolimits_{\mathop k\nolimits_l^u } ^H}\mathop {\mathbf{W}}\nolimits_{\mathop k\nolimits_l^u } .
\end{equation}		
	\indent	From \eqref{Al} and \eqref{Aukl} we can observe that all  $\{ {\mathop {\bf{A}}\nolimits_l ,\mathop {\bf{A}}\nolimits_{\mathop k\nolimits_l^u } ;\forall k,l} \}$ are  complex symmetric matrices, each of them  can be decomposed by the eigenvalue decomposition (ED) as ${\bf{\tilde A}} = {\bf{\tilde Q\tilde \Lambda }}{{{\bf{\tilde Q}}}^H}$, where ${{\bf{\tilde Q}}}$ is an orthogonal matrix satisfying  ${\bf{\tilde Q}}{{{\bf{\tilde Q}}}^H} = {{{\bf{\tilde Q}}}^H}{\bf{\tilde Q}} = {\bf{I}}$, and ${{\bf{\tilde \Lambda }}}$ is a diagonal matrix with non-negative elements. Therefore, the power constraints of $\{ {\mathop {\bf{F}}\nolimits_{\mathop k\nolimits_l^d } ,\mathop {\bf{F}}\nolimits_{\mathop k\nolimits_l^u } }\}$ can be reformulated as
		\begin{equation}\label{Lagral}
			\begin{aligned}
	{{{f}}_l}\left( {\mathop \lambda \nolimits_l } \right) &= \sum\limits_{k = 1}^{\mathop K\nolimits_l^d } {} {\text{Tr}}\left( {\mathop {\mathbf{F}}\nolimits_{\mathop k\nolimits_l^d }^H {{\left( {\mathop \lambda \nolimits_l } \right)}}\mathop {\mathbf{F}}\nolimits_{\mathop k\nolimits_l^d } \left( {\mathop \lambda \nolimits_l } \right)} \right) \hfill \\
	&= \sum\limits_{i = 1}^{M_B^t} {} \frac{{{{\left[ {{{\mathbf{Z}}_{{l}}}} \right]}_{i,i}}}}{{{{\left( {{{\left[ {{{\mathbf{\Lambda }}_l}} \right]}_{i,i}} + \mathop \lambda \nolimits_l } \right)}^2}}} \hfill ,
			\end{aligned}
		\end{equation}
			\begin{equation}\label{Lagraukl}
		\begin{aligned}
	\mathop {{f}}\nolimits_{\mathop k\nolimits_l^u } \left( {\mathop \lambda \nolimits_{\mathop k\nolimits_l^u } } \right) &= {\text{Tr}}\left( {\mathop {\mathbf{F}}\nolimits_{\mathop k\nolimits_l^u }^H {{\left( {\mathop \lambda \nolimits_{\mathop k\nolimits_l^u } } \right)}}\mathop {\mathbf{F}}\nolimits_{\mathop k\nolimits_l^u } \left( {\mathop \lambda \nolimits_{\mathop k\nolimits_l^u } } \right)} \right) \hfill \\
	&= \sum\limits_{i = 1}^{M_U^t} {} \frac{{{{\left[ {{{\mathbf{Z}}_{\mathop k\nolimits_l^u }}} \right]}_{i,i}}}}{{{{\left( {{{\left[ {{{\mathbf{\Lambda }}_{\mathop k\nolimits_l^u }}} \right]}_{i,i}} + \mathop \lambda \nolimits_{\mathop k\nolimits_l^u } } \right)}^2}}} \hfill ,
		\end{aligned}
	\end{equation}
where 
		\begin{equation}\label{}
{{\mathbf{Z}}_{{l}}} = \sum\limits_{k = 1}^{\mathop K\nolimits_l^d } {} {{\mathbf{Q}}_l^H}{\mathop {{\mathbf{\bar H}}}\nolimits_{\mathop k\nolimits_l^d ,l} ^H}{\mathop {\mathbf{U}}\nolimits_{\mathop k\nolimits_l^d } ^H}\mathop {\mathbf{W}}\nolimits_{\mathop k\nolimits_l^d } {\mathop {\mathbf{W}}\nolimits_{\mathop k\nolimits_l^d } ^H}\mathop {\mathbf{U}}\nolimits_{\mathop k\nolimits_l^d } \mathop {{\mathbf{\bar H}}}\nolimits_{\mathop k\nolimits_l^d ,l} {{\mathbf{Q}}_l},
		\end{equation}
			\begin{equation}\label{}
{{\mathbf{Z}}_{\mathop k\nolimits_l^u }} = {{\mathbf{Q}}_{\mathop k\nolimits_l^u }^H}{\mathop {{\mathbf{\bar H}}}\nolimits_{l,\mathop k\nolimits_l^u } ^H}{\mathop {\mathbf{U}}\nolimits_{\mathop k\nolimits_l^u } ^H}\mathop {\mathbf{W}}\nolimits_{\mathop k\nolimits_l^u } {\mathop {\mathbf{W}}\nolimits_{\mathop k\nolimits_l^u } ^H}\mathop {\mathbf{U}}\nolimits_{\mathop k\nolimits_l^u } \mathop {{\mathbf{\bar H}}}\nolimits_{l,\mathop k\nolimits_l^u } {{\mathbf{Q}}_{\mathop k\nolimits_l^u }}.
	\end{equation}
 It can be proved that (\ref{Lagral}) and (\ref{Lagraukl}) are  monotonically decreasing functions with respect to their corresponding Lagrangian multipliers, which can be solved through bisection search. For the determination of multipliers' upper bound, readers can refer to \cite{MulticellMIMOCommunications}. Here we omit the details due to the page limit. 
		\section{Phase Shift Matrix Optimization}
		In this section, we aim to obtain the optimal phase shift matrix ${\bf{\Phi }}$ in \eqref{OF:MMSE}. Specifically, substituting (\ref{Eqa:eB})-(\ref{Eqa:eU}) into (\ref{OF:MMSE}) and removing the constant and irrelevant terms, we obtain the phase shift optimization problem as follows
		
		\begin{small}
\begin{subequations}\label{OF:PHASE1}
	\begin{align}
	& \mathop {\min }\limits_{\bf{\Phi }} {\text{  }}\sum\limits_l^L {\sum\limits_k^{\mathop K\nolimits_l^u } {{\text{Tr}}\left( {\mathop {\bf{W}}\nolimits_{\mathop k\nolimits_l^u } \mathop {\bf{E}}\nolimits_{\mathop k\nolimits_l^u } } \right)} }  + \sum\limits_l^L {\sum\limits_k^{\mathop K\nolimits_l^d } {{\text{Tr}}\left( {\mathop {\bf{W}}\nolimits_{\mathop k\nolimits_l^d } \mathop {\bf{E}}\nolimits_{\mathop k\nolimits_l^d } } \right)} } {\text{ }}  \\
	& s.t. \quad {\text{   }}0 \leq \theta  \leq 2\pi ,\left| {\mathop \varphi \nolimits_m } \right| = 1,{\text{   }}\forall m ,
	\end{align}
\end{subequations}
		\end{small}\noindent
 By expanding the equivalent channel matrices and using the following equations \cite{zhang2017matrix}
			\begin{equation}\label{}
			{\rm{Tr}}\left( {{{\bf{\Phi }}^H}{\bf{D\Phi E}}} \right) = {\bm{\phi} ^H}\left( {{\bf{D}} \odot {{\bf{E}}^T}} \right)\bm{\phi},
		\end{equation}
		\begin{equation}\label{}
			{\rm{Tr}}\left( {{\bf{C}}{{\bf{\Phi }}^H}} \right){\rm{  = }}{\bm{\phi} ^H}{\bf{c}}{\rm{ , Tr}}\left( {{{\bf{C}}^H}{\bf{\Phi }}} \right){\rm{  = }}{{\bf{c}}^H}\bm{\phi}, 
		\end{equation}	
we reformulate (\ref{OF:PHASE1}) into a simpler form as follows
\begin{subequations}\label{OF:PHASE3}
		\begin{align}\label{OF:PHASE3a}
		&	\mathop {\min }\limits_{\bm{\phi }} f\left( \bm{\phi}  \right){\rm{  = }}{{\bf{c}}^H}\bm{\phi}  + {\bm{\phi} ^H}{\bf{c}} + {\bm{\phi} ^H}{\bf{\Xi }}\bm{\phi} \tag{36a} \\
		&	s.t. \quad \left| {\mathop \varphi \nolimits_m } \right| = 1,\forall m. \tag{36b}
	\end{align}
\end{subequations}
where ${\bf{c}} = {\left[ {{{\left[ {\bf{C}} \right]}_{1,1}},.....,{{\left[ {\bf{C}} \right]}_{M,M}}} \right]^T}$, ${\bf{C}}$ and ${\bf{\Xi }}$ are shown at the bottom of the next page. 

The problem (\ref{OF:PHASE3}) is still non-convex due to the unit modulus constraint on each component of  $\bm{\phi} $. Next, we propose a pair of methods to solve it, and compare their convergence and complexity performance in Section V.

		\newcounter{TempEqCnt5} 
		\setcounter{TempEqCnt5}{\value{equation}} 
		\setcounter{equation}{36} 
		\begin{figure*}[hb] 
			\hrulefill  
			\begin{equation}
				\begin{aligned}
	&{\bf{C}}{\rm{ = }}\sum\limits_{l = 1}^L {\sum\limits_{k = 1}^{\mathop K\nolimits_l^u } {} } \left( {\sum\limits_{j = 1}^L {\sum\limits_{i = 1}^{\mathop K\nolimits_j^d } {{\mathop {\bf{G}}\nolimits_{l,R} ^H}{\mathop {\bf{U}}\nolimits_{\mathop k\nolimits_l^u } ^H}\mathop {\bf{W}}\nolimits_{\mathop k\nolimits_l^u } \mathop {\bf{U}}\nolimits_{\mathop k\nolimits_l^u } \mathop {\bf{H}}\nolimits_{l,j} \mathop {\bf{F}}\nolimits_{\mathop i\nolimits_j^d } {\mathop {\bf{F}}\nolimits_{\mathop i\nolimits_j^d } ^H}{\mathop {\bf{G}}\nolimits_{R,j} ^H}} }  - {\mathop {\bf{G}}\nolimits_{l,R} ^H}{\mathop {\bf{U}}\nolimits_{\mathop k\nolimits_l^u } ^H}\mathop {\bf{W}}\nolimits_{\mathop k\nolimits_l^u } {\mathop {\bf{F}}\nolimits_{\mathop k\nolimits_l^u } ^H}{\mathop {\bf{G}}\nolimits_{R,\mathop k\nolimits_l^u } ^H}} \right.\\&
	\left. { + \sum\limits_{i = 1}^{\mathop K\nolimits_l^d } {} \left( {\mathop \rho \nolimits_{l,l}^{ - 1}  - 1} \right){\mathop {\bf{G}}\nolimits_{l,R} ^H}{\mathop {\bf{U}}\nolimits_{\mathop k\nolimits_l^u } ^H}\mathop {\bf{W}}\nolimits_{\mathop k\nolimits_l^u } \mathop {\bf{U}}\nolimits_{\mathop k\nolimits_l^u } \mathop {\bf{H}}\nolimits_{l,l} \mathop {\bf{F}}\nolimits_{\mathop i\nolimits_l^d } {\mathop {\bf{F}}\nolimits_{\mathop i\nolimits_l^d } ^H}{\mathop {\bf{G}}\nolimits_{R,l} ^H} + \sum\limits_{j = 1}^L {\sum\limits_{i = 1}^{\mathop K\nolimits_j^u } {{\mathop {\bf{G}}\nolimits_{l,R} ^H}{\mathop {\bf{U}}\nolimits_{\mathop k\nolimits_l^u } ^H}\mathop {\bf{W}}\nolimits_{\mathop k\nolimits_l^u } \mathop {\bf{U}}\nolimits_{\mathop k\nolimits_l^u } \mathop {\bf{H}}\nolimits_{l,\mathop i\nolimits_j^u } \mathop {\bf{F}}\nolimits_{\mathop i\nolimits_j^u } {\mathop {\bf{F}}\nolimits_{\mathop i\nolimits_j^u } ^H}{\mathop {\bf{G}}\nolimits_{R,\mathop i\nolimits_j^u } ^H}} } } \right)\\
	&+ \sum\limits_l^L {\sum\limits_k^{\mathop K\nolimits_l^d } {} } \left( {\sum\limits_{j = 1}^L {} \sum\limits_{i = 1}^{\mathop K\nolimits_j^d } {} {\mathop {\bf{G}}\nolimits_{\mathop k\nolimits_l^d ,R} ^H}{\mathop {\bf{U}}\nolimits_{\mathop k\nolimits_l^d } ^H}\mathop {\bf{W}}\nolimits_{\mathop k\nolimits_l^d } \mathop {\bf{U}}\nolimits_{\mathop k\nolimits_l^d } \mathop {\bf{H}}\nolimits_{\mathop k\nolimits_l^d ,j} \mathop {\bf{F}}\nolimits_{\mathop i\nolimits_j^d } {\mathop {\bf{F}}\nolimits_{\mathop i\nolimits_j^d } ^H}{\mathop {\bf{G}}\nolimits_{R,j} ^H}{\rm{ + }}\sum\limits_{j = 1}^L {\sum\limits_{i = 1}^{\mathop K\nolimits_j^u } {} } {\mathop {\bf{G}}\nolimits_{\mathop k\nolimits_l^d ,R} ^H}{\mathop {\bf{U}}\nolimits_{\mathop k\nolimits_l^d } ^H}\mathop {\bf{W}}\nolimits_{\mathop k\nolimits_l^d } \mathop {\bf{U}}\nolimits_{\mathop k\nolimits_l^d } \mathop {\bf{H}}\nolimits_{\mathop k\nolimits_l^d ,\mathop i\nolimits_j^u } \mathop {\bf{F}}\nolimits_{\mathop i\nolimits_j^u } {\mathop {\bf{F}}\nolimits_{\mathop i\nolimits_j^u } ^H}{\mathop {\bf{G}}\nolimits_{R,\mathop i\nolimits_j^u } ^H}} \right.\\&
	\left. { - {\rm{   }}{\mathop {\bf{G}}\nolimits_{\mathop k\nolimits_l^d ,R} ^H}{\mathop {\bf{U}}\nolimits_{\mathop k\nolimits_l^d } ^H}\mathop {\bf{W}}\nolimits_{\mathop k\nolimits_l^d } {\mathop {\bf{F}}\nolimits_{\mathop k\nolimits_l^d } ^H}{\mathop {\bf{G}}\nolimits_{R,l} ^H}} \right),
				\end{aligned}
			\end{equation}
			\begin{equation}
	\begin{aligned}
	&{\bf{\Xi }} = \sum\limits_{l = 1}^L {} \left( {\sum\limits_{k = 1}^{\mathop K\nolimits_l^u } {} \left( {\mathop \rho \nolimits_{l,l}^{ - 1}  - 1} \right){\mathop {\bf{G}}\nolimits_{l,R} ^H}{\mathop {\bf{U}}\nolimits_{\mathop k\nolimits_l^u } ^H}\mathop {\bf{W}}\nolimits_{\mathop k\nolimits_l^u } \mathop {\bf{U}}\nolimits_{\mathop k\nolimits_l^u } \mathop {\bf{G}}\nolimits_{l,R} } \right) \odot {\left( {\sum\limits_{i = 1}^{\mathop K\nolimits_l^d } {} \mathop {\bf{G}}\nolimits_{R,l} \mathop {\bf{F}}\nolimits_{\mathop i\nolimits_l^d } {\mathop {\bf{F}}\nolimits_{\mathop i\nolimits_l^d } ^H}{\mathop {\bf{G}}\nolimits_{R,l} ^H}} \right)^T}\\
	&+ \left( {\sum\limits_l^L {\sum\limits_k^{\mathop K\nolimits_l^d } {} } {\mathop {\bf{G}}\nolimits_{\mathop k\nolimits_l^d ,R} ^H}{\mathop {\bf{U}}\nolimits_{\mathop k\nolimits_l^d } ^H}\mathop {\bf{W}}\nolimits_{\mathop k\nolimits_l^d } \mathop {\bf{U}}\nolimits_{\mathop k\nolimits_l^d } \mathop {\bf{G}}\nolimits_{\mathop k\nolimits_l^d ,R}  + \sum\limits_{l = 1}^L {\sum\limits_{k = 1}^{\mathop K\nolimits_l^u } {{\mathop {\bf{G}}\nolimits_{l,R} ^H}{\mathop {\bf{U}}\nolimits_{\mathop k\nolimits_l^u } ^H}\mathop {\bf{W}}\nolimits_{\mathop k\nolimits_l^u } \mathop {\bf{U}}\nolimits_{\mathop k\nolimits_l^u } \mathop {\bf{G}}\nolimits_{l,R} } } } \right) \\& \odot {\left( {\sum\limits_{j = 1}^L {} \sum\limits_{i = 1}^{\mathop K\nolimits_j^d } {} \mathop {\bf{G}}\nolimits_{R,j} \mathop {\bf{F}}\nolimits_{\mathop i\nolimits_j^d } {\mathop {\bf{F}}\nolimits_{\mathop i\nolimits_j^d } ^H}{\mathop {\bf{G}}\nolimits_{R,j} ^H} + \sum\limits_{j = 1}^L {\sum\limits_{i = 1}^{\mathop K\nolimits_j^u } {} } \mathop {\bf{G}}\nolimits_{R,\mathop i\nolimits_j^u } \mathop {\bf{F}}\nolimits_{\mathop i\nolimits_j^u } {\mathop {\bf{F}}\nolimits_{\mathop i\nolimits_j^u } ^H}{\mathop {\bf{G}}\nolimits_{R,\mathop i\nolimits_j^u } ^H}} \right)^T},
	\end{aligned}
\end{equation}		
		\end{figure*}

		\subsection{Complex Circle Manifold (CCM) Algorithm Development}
		We find that the phase shift vector $\bm{\phi} $ can be regarded as an $M$-dimensional complex circle manifold 
	\begin{equation}\label{SM}
{\mathcal{S}^M} = \left\{ {\bm{\phi}  \in {\mathbb{C}^M}:{\text{ }}\left| {{\varphi _m}} \right| = 1,m = 1,2, \cdots, M} \right\}.
	\end{equation}
Hence, the tangent space at the point $\bm\phi  \in {\mathcal{S}^M}$ is
\begin{equation}\label{TM}
{\mathcal{T}_{\bm\phi} }{\mathcal{S}^M} = \left\{ {{\mathbf{z}} \in {\mathbb{C}^M}: {\operatorname{Re} \left\{ {{\mathbf{z}} \odot \bm{\phi} } \right\}}  = {{\mathbf{0}}_M}} \right\}.
\end{equation}
Similar to the gradient descent method used in Euclidean space, the main idea of CCM is to find the gradient in a manifold space and then descending along with the negative gradient direction. Typically, there are four main steps of the CCM algorithm.

\IncMargin{1em} 

\begin{algorithm}[t]
	\SetKwInOut{Input}{\textbf{Input}}
	\SetKwInOut{Output}{\textbf{Output}} 
	
	
	Set $t{\text{ = 0}}$, initialize the maximum number of iterations ${t_{CCM }}$ and error tolerance $\varepsilon_{CCM}$, randomly generate ${\bm{\phi} ^{\text{0}}}$ and calculate the value of function (\ref{OF:PHASE3a}) as $ f\left( {{\bm{\phi} ^0}} \right)$\;
	\Repeat
	{$\left| {\frac{{ f\left( \bm{\phi}^{t{\rm{ }}}  \right){\rm{  - }} f\left( {{\bm{\phi} ^{t{\rm{ + 1}}}}} \right)}}{{ f\left( {{\bm{\phi} ^{t{\rm{ + 1}}}}} \right)}}{\rm{ }}} \right| \le \varepsilon_{CCM} $ or ${\text{t}} \ge {{\text{t}}_{CCM}}$ } 
	{Calculate the Euclidean gradient ${{\bm{\eta }}^t}$ in (\ref{Eq:Euclidean Gradient})\;
		Calculate the Riemannian gradient ${P_{{\mathcal{T}_{{{\bm{\phi}}^t}}}{\mathcal{S}^M}}}\left( {{{\bm{\eta }}^t}} \right)$ in (\ref{Eq:Riemanian Gradient})\;
		Calculate step size $\zeta^t $ according to the Armijo criterion\;
		Update over the tangent space according to (\ref{Eq:Update over the tangent space})\;
		Calculate ${\bm{\phi} ^{t{\text{ + 1}}}}$ by retraction ${\bm{\phi} ^t}$ to the complex circle manifold ${{\mathcal{S}^M}}$ according to (\ref{Eq:Retraction operator})\;
		Calculate the objective function ${ f\left( {{\bm{\phi} ^{t{\text{ + 1}}}}} \right)}$\;
		Update $t \leftarrow t{\text{ + 1}}$\;
	}
	Return $\bm{\phi} ={\bm{\phi} ^{t{\text{ + 1}}}}$.
	\caption{CCM Algorithm}\label{CCM}
\end{algorithm}

		\subsubsection{Euclidean Gradient}
		The gradient in a complex Euclidean space is defined as
		\begin{equation}\label{}
			\nabla f\left( x \right) = 2\frac{{\partial f\left( x \right)}}{{\partial {x^*}}}.
		\end{equation}
Hence, the negative Euclidean gradient of the problem (\ref{OF:PHASE3}) in the $t$th iteration is given by
		\begin{equation}\label{Eq:Euclidean Gradient}
{{\bm{\eta }}^t} =  - \nabla {\text{ }} f\left( {{\bm{\phi} ^t}} \right) =  - {\text{2}}{\mathbf{\Xi }}{\bm{\phi} ^t} - {\text{2}}{\mathbf{c}}.
		\end{equation}
		\subsubsection{Riemannian Gradient}
		By using the orthogonal projection operator, we can project the Euclidean gradient at point ${{\bm{\phi} ^t}}$ onto the tangent space ${{\mathcal{T}_{{{\bm{\phi^t}}}}}{\mathcal{S}^M}}$, which is the space of the Riemannian gradient at point ${{\bm{\phi} ^t}}$. Explicitly, it is calculated as
		\begin{equation}\label{Eq:Riemanian Gradient}
{P_{{\mathcal{T}_{{{\bm{\phi}}^t}}}{\mathcal{S}^M}}}\left( {{{\bm{\eta }}^t}} \right)={{\bm{\eta }}^t} - \operatorname{Re} \left\{ {{{\bm{\eta }}^t}^* \odot {\bm{\phi} ^t}} \right\}  \odot {\bm{\phi} ^t}.
		\end{equation}
	
		\subsubsection{Update over the Tangent Space}
		By applying certain line search method, we obtain the updated point ${{\bm{\phi} ^t}}$ on the tangent space ${{{\mathcal{T}_{{{\bm{\phi}}^t}}}{\mathcal{S}^M}}}$ as follows
\begin{equation}\label{Eq:Update over the tangent space}
{{{\bm{\bar \phi }}}^t}={{{\bm{ \phi }}}^t}{\text{ + }}\zeta^t {P_{{\mathcal{T}_{{{\bm{\phi}}^t}}}{\mathcal{S}^M}}}\left( {{{\bm{\eta }}^t}} \right),
		\end{equation}
		where $\zeta^t  $ is the step size in the $t$th iteration. In this paper, we utilize the exact line search, which would provide fine resolution. Specifically, $\zeta^t$ is determined according to the Armijo criterion, which should satisfy the following criterion
			\begin{equation}
f\left( {{\bm{\bar \phi }^t}} \right) - f\left( {\bm{\phi ^t}} \right) \le  - \tau \zeta^t {P_{{\mathcal{T}_{\bm{\phi ^t}}}{\mathcal{S}^M}}}{\left( {{{\bm{\eta }}^t}} \right)^H}{P_{{\mathcal{T}_{\bm{\phi ^t}}}{\mathcal{S}^M}}}\left( {{{\bm{\eta }}^t}} \right).
		\end{equation}
 The parameters above should be chosen to satisfy $\tau  \in \left( {{\text{0,0}}{\text{.5}}} \right)$	and $\zeta^t  {\rm{ > }} {\text{0}}$ \cite{WeightedSumRateMaximization}.
		\subsubsection{Retraction Operator}
		Since the updated point ${{{\bm{\bar \phi }}}^t}$ is not in ${\mathcal{S}^M}$, we apply a retraction operator to map ${{{\bm{\bar \phi }}}^t}$ into the complex circle manifold as follows
		\begin{equation}\label{Eq:Retraction operator}
			{\bm{\phi} ^{t{\rm{ + 1}}}}{\rm{ = }}{{{\bm{\bar \phi }}}^t} \odot \frac{{\rm{1}}}{{\left| {{{\bm{\bar \phi }}}^t} \right|}}.
		\end{equation}
\indent With above steps, we can obtain a suboptimal solution to the problem (\ref{OF:PHASE3}). The details of CCM algorithm are shown in Algorithm \ref{CCM}.

		\subsection{Successive Convex Approximation (SCA) Algorithm Development}
In order to meet the unit modulus constraint, the aforementioned CCM algorithm calculates the Euclidean gradient to obtain the Riemannian one, and then retract the updated points back into the manifold, which is relatively twisting. As an alternative, we can deal with the unit modulus in a more straightforward way. Particularly, since $\phi_m= {{{\text{e}}^{j{\theta _m}}}}$ is defined and ${{\theta _m}}$ is a continuous variable, we can solve the problem (\ref{OF:PHASE3}) based on ${{\bm{\theta}}}$, where ${\bm{\theta }} = {\left[ {{\theta _1},...,{\theta _M}} \right]^T}$.  Inspired by this, we rewrite (\ref{OF:PHASE3}) into an equivalent form as
\begin{subequations}\label{OF:SCA}
		\begin{align}\label{OF:SCAa}
		&\mathop {\min }\limits_{\bm{\theta }} {\text{   }}f\left( {\bm{\theta }} \right){\text{  = }}{{\mathbf{c}}^H}{e^{j{\bm{\theta }}}} + {\left( {{e^{j{\bm{\theta }}}}} \right)^H}{\mathbf{c}}{\text{ + }}{\left( {{e^{j{\bm{\theta }}}}} \right)^H}{\bm{\Xi }}{e^{j{\bm{\theta }}}} \hfill \tag{47a} \\
		&	s.t. \quad 0 \leq {\theta _m} \leq 2\pi , \forall m  \hfill \tag{47b} .
	\end{align}
\end{subequations}
Notice that the feasible set of the problem \eqref{OF:SCA} is an $M$-dimensional $\left[0,2 \pi\right)$ set, which is convex and denoted as $\mathbb{Q}$. However, it is challenging to obtain the optimal solution of  (\ref{OF:SCA}) directly. Here we turn to the SCA method by constructing a series of convex problems and successively approximating to a suboptimal solution. In the following, we show the main idea of the SCA technique. 
 
 {Instead of minimizing the function $f\left( {\bm{\theta }} \right)$ directly, the algorithm optimizes a series of surrogate functions  that minimize  $f\left( {\bm{\theta }} \right)$. Here we denote the surrogate function for $f\left( {{{\bm{\theta }}^{ t }}} \right)$ by $\bar f\left( {{\bm{\theta }},{{\bm{\theta }}^{ t }}} \right)$. More specifically, starting from a feasible point ${{\bm{\theta }}^{ 0 }}$, the algorithm generates a series of $\left\{ {{{\bm{\theta }}^{ t }}} \right\}$ according to the following update rule}
		\begin{equation}\label{Eq:surrogate}
	{{\bm{\theta }}^{ {t + 1} }} = \mathop {\arg  }\min\limits_{{{\bm{\theta }}\in {\mathbb{Q}} }} {\text{ }}\bar f\left( {{\bm{\theta }},{{\bm{\theta }}^{ t }}} \right)
\end{equation}				
where ${{{\bm{\theta }}^{ t }}}$ is the point generated by the algorithm in the $(t-1)$th iteration. The problem (\ref{Eq:surrogate}) is referred to as the surrogate problem in the $t$th iteration. The iteration will stop until certain criterion is met.  Moreover, to guarantee the convergence of the algorithm, the surrogate function $\bar f\left( {{\bm{\theta }},{{\bm{\theta }}^{ t }}} \right)$ should satisfy the following four conditions
 \begin{subequations}\label{Eq:SCAcondition}
\begin{align}
	\label{Eq:SCAconditiona}
	&\bar f\left( {{\bm{\theta }},{{\bm{\theta }}^{ t }}} \right){\text{is continuous in }}{\bm{\theta }}{\text{ and }}{{\bm{\theta }}^{ t }}\tag{49a},\\ 
	\label{Eq:SCAconditionb}
	&{\left. {\bar f'\left( {{\bm{\theta }},{{\bm{\theta }}^{ t }};\bm{d}} \right)} \right|_{{\bm{\theta }} = {{\bm{\theta }}^{ t }}}} = f'\left( {{{\bm{\theta }}^{ t }};\bm{d}} \right),\forall \bm{d}{\text{ with }}{{\bm{\theta }}^{ t }} + \bm{d} \in {\Bbb Q}\tag{49b},\\
	\label{Eq:SCAconditionc}
	&\bar f\left( {{{\bm{\theta }}^{ t }},{{\bm{\theta }}^{ t }}} \right) = f\left( {{{\bm{\theta }}^{ t }}} \right),\forall {{\bm{\theta }}^{ t }} \in {\Bbb Q},\tag{49c}\\
	\label{Eq:SCAconditiond}
	&\bar f\left( {{\bm{\theta }},{{\bm{\theta }}^{ t }}} \right) \ge f\left( {\bm{\theta }} \right), \forall {\bm{\theta }},{{\bm{\theta }}^{ t }} \in {\Bbb Q}\tag{49d}.
\end{align}	
 \end{subequations} 
 \noindent Condition (\ref{Eq:SCAconditionb}) ensures that the derivatives of the surrogate function $\bar f\left( {{\bm{\theta }},{{\bm{\theta }}^{ t }}} \right)$ and the original function $f\left( {\bm{\theta }} \right)$ must be equivalent at point ${{{\bm{\theta }}^{ t }}}$. Conditions (\ref{Eq:SCAconditionc}) and (\ref{Eq:SCAconditiond}) imply that the surrogate function is a tight upper bound of the original function. As a result, (\ref{Eq:SCAcondition}) ensures that (\ref{Eq:surrogate}) converges to a stationary point, which is a suboptimal solution to \eqref{OF:SCA} \cite[Theorem 1]{Aunifiedconvergenceanalysis}.

It is obvious that (\ref{OF:SCAa}) is continuously differentiable. Therefore, by performing a second-order Taylor expansion of  (\ref{OF:SCAa}), we obtain a surrogate function as follows \cite{WeightedSumRateMaximization}
	\begin{equation}\label{OF:sur}
		\begin{aligned}
\bar f\left( {{\bm{\theta }},{{\bm{\theta }}^{ t }}} \right)=&f\left( {{{\bm{\theta }}^{ t }}} \right) + \nabla f{\left( {{{\bm{\theta }}^{ t }}} \right)^T}\left( {{\bm{\theta }} - {{\bm{\theta }}^{ t }}} \right) \\&+ \frac{\beta }{2}\left\| {{\bm{\theta }} - {{\bm{\theta }}^{ t }}} \right\|_2^2
\end{aligned}
\end{equation}	
where $\nabla f\left( {{{{\bm{\theta }}^{ t }}}} \right)$ is the gradient of (\ref{OF:SCAa}) and  $\nabla f\left( {\bm{\theta }} \right) = 2\operatorname{Re} \left\{ { - j{{{\bm{ \phi }}}^*} \odot  \left( {{\bm{\Xi  \phi }} + {\mathbf{c}}} \right)} \right\}$. It is easy to prove that when parameters ${\mathbf{c}}$  and  ${\mathbf{\Xi }}$ are determined,  the four conditions in (\ref{Eq:SCAcondition}) are satisfied with a sufficiently large $\beta $. Finally, by taking the first derivative of  (\ref{OF:sur}) w.r.t. ${\bm{\theta }}$ and setting it to zeros, the optimal ${{\bm{\theta }}^{ {t + 1} }}$ could be obtained as follows
\begin{equation}\label{Eq:theta_update}
{{\bm{\theta }}^{ {t + 1} }}={{\bm{\theta }}^{ t }}-\frac{{\nabla f\left( {{{\bm{\theta }}^{ t }}} \right)}}{\beta }.
\end{equation}
We also apply the Armijo criterion as shown in (\ref{Eq:Update over the tangent space}) to determine a suitable $\beta $. The steps of the SCA algorithm are listed in Algorithm \ref{SCA}.

\DecMargin{1em}
		\IncMargin{1em} 
		\begin{algorithm}[t]
			\SetKwInOut{Input}{\textbf{Input}}
			\SetKwInOut{Output}{\textbf{Output}} 
			
			
			Set $t{\text{ = 0}}$, initialize the maximum number of iterations ${t_{SCA }}$ and error tolerance $\varepsilon_{SCA}$, randomly generate ${\bm{\theta} ^{\text{0}}}$ and calculate the value of function (\ref{OF:SCAa}) as $ f\left( {{\bm{\theta} ^0}} \right)$\;
			\Repeat
			{$\left| {\frac{{ f\left( \bm{\theta}^{t{\rm{ }}}  \right){\rm{  - }} f\left( {{\bm{\theta} ^{t{\rm{ + 1}}}}} \right)}}{{ f\left( {{\bm{\theta} ^{t{\rm{ + 1}}}}} \right)}}{\rm{ }}} \right| \le \varepsilon_{SCA} $ or ${\text{t}} \ge {{\text{t}}_{SCA}}$}
			{Calculate the Euclidean gradient $\nabla f\left( {\bm{\theta }^{t{\rm{}}}} \right) $\;
				Calculate $\beta^{t} $ according to Armijo criterion:\;
				Update ${\bm{\theta} ^{t{\text{ + 1}}}}$ according to (\ref{Eq:theta_update})\;
				Calculate the objective function ${ f\left( {{\bm{\theta} ^{t{\text{ + 1}}}}} \right)}$\;
				Update $t \leftarrow t{\text{ + 1}}$\;
			}
			Return $\bm{\theta} ={\bm{\theta} ^{t{\text{ + 1}}}}$.
			\caption{SCA Algorithm}\label{SCA}
		\end{algorithm}
		\DecMargin{1em}
		
		\subsection{Overall Algorithm to Solve Problem (\ref{OF:original})}	
		Based on above analysis, the overall BCD-based method proposed to solve \eqref{OF:original} is summarized in Algorithm \ref{BCD-CCM}.  In each iteration, we optimize ${\mathbf{F}}$ and ${\mathbf{\Phi }}$ in an alternating way while fixing ${\bf{U}}$ and ${\bf{W}}$.	
		When the phase shift matrix ${\mathbf{\Phi }}$ is fixed,  the Lagrangian multiplier method and bisection search algorithm ensure the optimal overall TPC matrices ${\mathbf{F}}$ of (\ref{OF:original}). Then, by fixing ${\mathbf{F}}$ and optimizing ${\mathbf{\Phi }}$ through the CCM or SCA algorithm, we obtain its suboptimal solution. The two groups of variables are alternately updated until the OF of the problem	(\ref{OF:original}) converges.   
\IncMargin{1em} 
\begin{algorithm}[t]
	\SetKwInOut{Input}{\textbf{Input}}
	\SetKwInOut{Output}{\textbf{Output}} 
	
	\Input{$\mathop {\bf{H}}\nolimits_{\mathop k\nolimits_l^d ,j} {\rm{ , }}\mathop {\bf{H}}\nolimits_{\mathop k\nolimits_l^d ,\mathop i\nolimits_j^u } {\rm{ , }}\mathop {\bf{H}}\nolimits_{l,j} {\rm{ , }}\mathop {\bf{H}}\nolimits_{l,\mathop i\nolimits_j^u }   $ \\
	$\mathop {\bf{G}}\nolimits_{l,R} {\rm{ , }}\mathop {\bf{G}}\nolimits_{R,l} {\rm{ , }}\mathop {\bf{G}}\nolimits_{R,\mathop k\nolimits_l^u {\rm{ }}} {\rm{ , }}\mathop {\bf{G}}\nolimits_{\mathop k\nolimits_l^d ,R}    $ \\  }
	\Output{${\mathbf{F}}, {\mathbf{\Phi }}$}
	\BlankLine   
	
	Randomly initialize the feasible solution ${{\mathbf{F}}^{0}}$ and  ${{\mathbf{\Phi }}^{0}}$ , calculate the value of function (\ref{OF:originala}) as ${\text{Obj}}({{\mathbf{F}}^{0}}{\mathbf{,}}{{\mathbf{\Phi }}^{0}})$ ;Set $t{\text{ = 0}}$ , the maximum number of iterations ${t_{\max }}$ and the error tolerance $\varepsilon$\;

	\Repeat
	{$t > {t_{\max }}$ or \\
	 $\left| {\frac{{{\rm{Obj}}({{\bf{F}}^{t + 1}}{\bf{,}}{{\bf{\Phi }}^{t + 1}}) - {\rm{Obj}}({{\bf{F}}^{t}}{\bf{,}}{{\bf{\Phi }}^{t}})}}{{{\rm{Obj}}({{\bf{F}}^{t}}{\bf{,}}{{\bf{\Phi }}^{t}})}}} \right| \le \varepsilon  $}
	{With ${{\bf{F}}^{t}}$ and ${{\bf{\Phi }}^{t}}$, compute the optimal decoding matrices ${{\bf{U}}^{t}}$ according to (\ref{Eqa:miuU}) and (\ref{Eqa:miuB}), and then compute the optimal weighting matrices ${{\bf{W}}^{t}}$ according to (\ref{Eqa:wB,wU})\;
	With ${{\bf{U}}^{t}}$ and ${{\bf{W}}^{t}}$, compute the optimal precoding matrices ${{\bf{F}}^{t + 1}}$ by solving problem (\ref{OF:TPC}) with the Largangian multiplier method and bisection search\;
With ${{\bf{U}}^{t}}$, ${{\bf{W}}^{t}}$ and ${{\bf{F}}^{t + 1}}$, compute the optimal ${{\bf{\Phi }}^{t + 1}}$ by solving problem (\ref{OF:PHASE1}) with Algorithm \ref{CCM} or Algorithm \ref{SCA}\;
Calculate the OF value of (\ref{OF:originala}) as ${{\text{Obj}}({{\mathbf{F}}^{t + 1}}{\mathbf{,}}{{\mathbf{\Phi }}^{t + 1}})}$ and let $t \leftarrow t{\text{ + 1}}$\;}

Return ${\mathbf{F}} = {{\mathbf{F}}^{t + 1}}$ and ${\mathbf{\Phi }} = {{\mathbf{\Phi }}^{t + 1}}$.
	\caption{The overall BCD-based Algorithm }\label{BCD-CCM}
\end{algorithm}
\DecMargin{1em}
		\subsection{Complexity Analysis}
	In this part, we analyze the complexity of the overall algorithm. Let us shed light on the complexity of the overall TPC matrices optimization firstly. In each iteration, the main complexity lies in updating ${\mathbf{F}}$ according to \eqref{UpdateFdkl} and \eqref{UpdateFukl}, which is $\mathcal{O}\left( L\left( {\mathop K\nolimits_{}^d M{{_B^t}^3}{\text{  + }}\mathop K\nolimits_{}^u M{{_U^t}^3}} \right) \right)$. The number of iterations to achieve convergence is determined by the bisection search algorithm, which is ${{{\log }_2}\left( {\frac{{{\lambda _u} - {\lambda _l}}}{{{\varepsilon _{bi}}}}} \right)}$, where ${{\varepsilon _{bi}}}$ , ${{\lambda _u}}$ and ${{\lambda _l}}$ are the error tolerance, the upper bound and lower bound of the bisection search algorithm, respectively. Hence, the overall complexity of calculating ${\mathbf{F}}$ is on the order of $\mathcal{O}\left( {{{\log }_2}\left( {\frac{{{\lambda _u} - {\lambda _l}}}{{{\varepsilon _{bi}}}}} \right)L\left( {\mathop K\nolimits_{}^d M{{_B^t}^3}{\rm{  + }}\mathop K\nolimits_{}^u M{{_U^t}^3}} \right)} \right)$.   
	
	As for the phase shift optimization, the complexity of calculating ${\mathbf{C}}$ and ${\mathbf{\Xi }}$ is ${\cal O}\left( {{L^2}\mathop K\nolimits_{}^d \mathop K\nolimits_{}^u {M^2}} \right)$, when the number of reflecting elements $M$ is much larger compared to the antenna size at BSs and UEs. For the major four steps of the CCM algorithm, the complexity depends on the calculation of the Euclidean gradient ${{{\bm{\eta }}^t}}$, which is given by ${\text{ }}\mathcal{O}\left( {{M^2}} \right)$. Denoting ${{T_{CCM}}}$ as the iteration number of the CCM algorithm, then the total complexity is  $\mathcal{O}\left( {{T_{CCM}}{M^2}} \right)$. Similarly, the complexity of the SCA algorithm in each iteration mainly depends on the computation of the Euclidean gradient and we denote $T_{SCA}$ as the related iteration number. Denoting the number of the outer BCD iteration is ${T}$, we have the overall complexity at the order of \normalsize{$\mathcal{O}\left( {T\left( {{{\log }_2}\left( {\frac{{{\lambda _u} - {\lambda _u}}}{{{\varepsilon _{bi}}}}} \right)L\left( {\mathop K\nolimits_{}^d M{{_B^t}^3}{\text{  + }}\mathop K\nolimits_{}^u M{{_U^t}^3}} \right) + {T_{CCM}}{M^2}} \right.} \right.$} $\left. {\left. { + {L^2}\mathop K\nolimits_{}^d \mathop K\nolimits_{}^u {M^2}} \right)} \right)$  and   \normalsize{$\mathcal{O}\left( {T\left( {{L^2}\mathop K\nolimits_{}^d \mathop K\nolimits_{}^u {M^2}} \right. + {T_{SCA}}{M^2}} \right.$} $ + \left. {\left. {{{\log }_2}\left( {\frac{{{\lambda _u} - {\lambda _u}}}{{{\varepsilon _{bi}}}}} \right)L\left( {\mathop K\nolimits_{}^d M{{_B^t}^3}{\text{  + }}\mathop K\nolimits_{}^u M{{_U^t}^3}} \right)} \right)} \right)$ \normalsize{for CCM and SCA algorithm, respectively.} 
		\section{Simulations and Discussions}
In this section, simulation results are provided for evaluating the performance of the proposed RIS-empowered FD networking design. 
		\subsection{Simulation Settings}
\begin{figure}[t]
 	\centering
 	\includegraphics[scale=0.2]{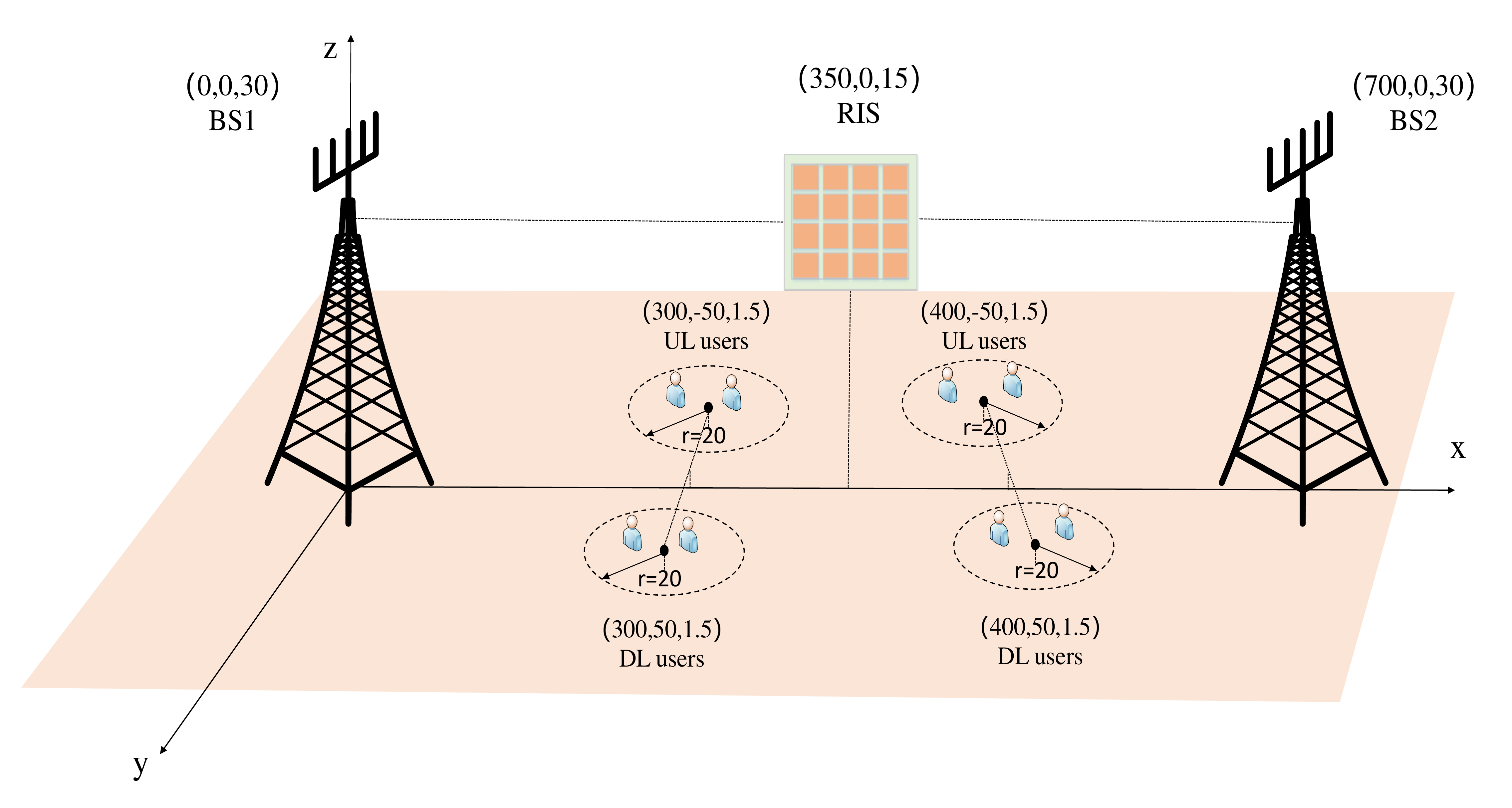}
 	\caption{Simulation scenario.}
 	\label{fig:SystemModel}
 \end{figure}

The system parameters are summarized as follows. The 3D-coordinate illustration to the simulation deployment is given in Fig. \ref{fig:SystemModel}, where we set the number of cells to $L = 2$. The coordinates of BS 1, RIS and BS 2 are  $(0,0,30)$, $(350,0,15)$, and $(700,0,30)$ meter, respectively. The DL and UL users in the first and the second cell are randomly and uniformly distributed in a circular region centered at ${{(x_{\rm{user}}}}, \pm {{y}_{\rm{user}}}, {\rm{1.5}})$ and $(400-x_{\rm{user}}, \pm {{y}_{\rm{user}}}, {\rm{1.5)}}$ with a radius of $20$ meter, respectively. We set $x_{\rm{user}}=300$ and $y_{\rm{user}}=50$ unless otherwise stated.
Notice that such user deployment is nearly on the cell edge. As depicted in Fig. \ref{fig:SystemModel}, the $z$-axis coordinate of BSs, RIS, and the users indicates their heights, which are $30$, $15$, and $1.5$ meter, respectively. Unless otherwise stated, the other parameters are set as follows: carrier frequency ${f_c} = 2.4$ GHz, bandwidth $B=10$ MHz, noise power density $\sigma _0^{\text{2}}{\text{ =}  - {174 }{\text{ }}{\text{dBm/Hz}}}$, number of users $\mathop K\nolimits_{}^d  = \mathop K\nolimits_{}^u {\rm{ = 2 }}$, number of the BS and user transmit antennas $\mathop M\nolimits_B^t  = \mathop M\nolimits_U^t  = 6$ , number of the BS and user receive  antennas $\mathop M\nolimits_B^r  = \mathop M\nolimits_U^r  = 2$, data streams $\mathop b\nolimits_d  = \mathop b\nolimits_u  = 2$, number of the reflecting elements $M=100$, SIC coefficient ${\rho _{{{l,l}}}} =90$~dB, error tolerance ${\varepsilon _{CCM}}{\text{ =  }}{\varepsilon _{SCA}} = {10^{ - 5}}$. Finally, the power budgets at BSs and users are $1$ W and $200$ mW, respectively.
 
		We generate the wireless channels in each realization according to the following description. For the large-scale fading, it is given by $PL(d) = {\rho _0} - 10\alpha {\lg}( {\frac{d}{{{d_0}}}} )$, where ${\rho _0} =  - 30$ dB is the path loss at the reference distance \cite{MulticellMIMOCommunications}, 
		$d$ is the Euclidean distance between two terminals, and $\alpha>0$ represents the path loss exponent. 
		
		As for the small-scale modeling, the Rayleigh fading is assumed for all BS-user links, while the Rician model is considered for the BS-RIS, RIS-user, BS-BS, and the SI links. Typically, the small-scale channel of Rician fading is given by
		\begin{equation}\label{}
		{\bf{\tilde H}} = \sqrt {\frac{\kappa  }{{\kappa  + 1}}} {{{\bf{\tilde H}}}^{LoS}} + \sqrt {\frac{1}{{\kappa  + 1}}} {{{\bf{\tilde H}}}^{NLoS}},
		\end{equation}
		where $\kappa $ is the Rician factor, ${{{\bf{\tilde H}}}^{LoS}}$ denotes the deterministic line of sight (LoS) or any dominant component, and ${{{\bf{\tilde H}}}^{NLoS}}$ denotes the non-LoS (NLoS) component that obeys Rayleigh fading. The LoS component is given by $
			{{{\bf{\tilde H}}}^{LoS}} = {{\bf{a}}_{Dr}}\left( \vartheta   \right){\bf{a}}_{Dt}^H\left( \psi   \right)
$
		with 
		\begin{equation}\label{}
			{{\bf{a}}_{Dr}}\left( \vartheta  \right) = {\left[ {1,{e^{j\frac{{2\pi }}{\lambda }{d_a }\sin \left( \vartheta  \right)}},...,{e^{j\frac{{2\pi }}{\lambda }(Dr - 1){d_a }\sin \left( \vartheta  \right)}}} \right]^T},
		\end{equation}
		\begin{equation}\label{}
			{{\bf{a}}_{Dt}}\left( \psi  \right) = {\left[ {1,{e^{j\frac{{2\pi }}{\lambda }{d_a }\sin \left( \psi  \right)}},...,{e^{j\frac{{2\pi }}{\lambda }(Dt - 1){d_a }\sin \left( \psi  \right)}}} \right]^T}.
		\end{equation}where $\vartheta $ and $\psi $ are the angle of arrival and departure (AoA/AoD), respectively. It is assumed that AoA and AoD are randomly distributed within $\left[ {0,2\pi } \right)$. ${Dr}$ and ${Dt}$ are the numbers of antennas at the receiver side and transmitter side, respectively. $\lambda $ is the signal wavelength, and ${d_a }$ is the spacing between adjacent antennas. In our simulations, we set antenna spacing ${d_a }= \lambda /2$. The path loss for the direct link of BS-user, user-user and BS-BS, the reflected link of BS-RIS, RIS-user are set to be ${\alpha _{B,U}} = 3.75$ , ${\alpha _{U,U}} = 3.9$, ${\alpha _{B,B}}=3.2$  and ${\alpha _{B,R}} = {\alpha _{U,R}}  \triangleq  {\alpha _R} = 2.2$, respectively. Since the transmit and receive antennas are deployed closely at FD BSs, we assume that the SI link does not suffer from the large-scale fading, i.e., its path loss is $0$ dB. The Rician factors are all set to be $3$. All simulation results are obtained by averaging over $400$ independent realizations.\\
\indent To investigate the performance gained by the proposed design, we include the following benchmark schemes:
	\subsubsection{FD Networking without RIS}	
		All phases of reflecting elements are set to be zero and the TPC matrices are optimized according to the method in Section III, i.e., CCM or SCA.
	\subsubsection{FD Networking with Random RIS}	
	A randomly generated phase shift matrix is applied to RIS. The TPC matrices at FD BSs and HD users are optimized according to Section III.
	\subsubsection{HD Networking with Optimized RIS}	
The BSs and users are all working in the HD mode. The TPC matrices at HD BSs and users, and the phase shifts at RIS are optimized according to Algorithm \ref{BCD-CCM} with SCA applied.
	\subsubsection{HD Networking without RIS}
	Multiple HD BSs form a mobile networking system, where the TPC matrices  are optimized according to the method in Section III.
		\begin{figure}[t]
	\centering
	\includegraphics[scale=0.3]{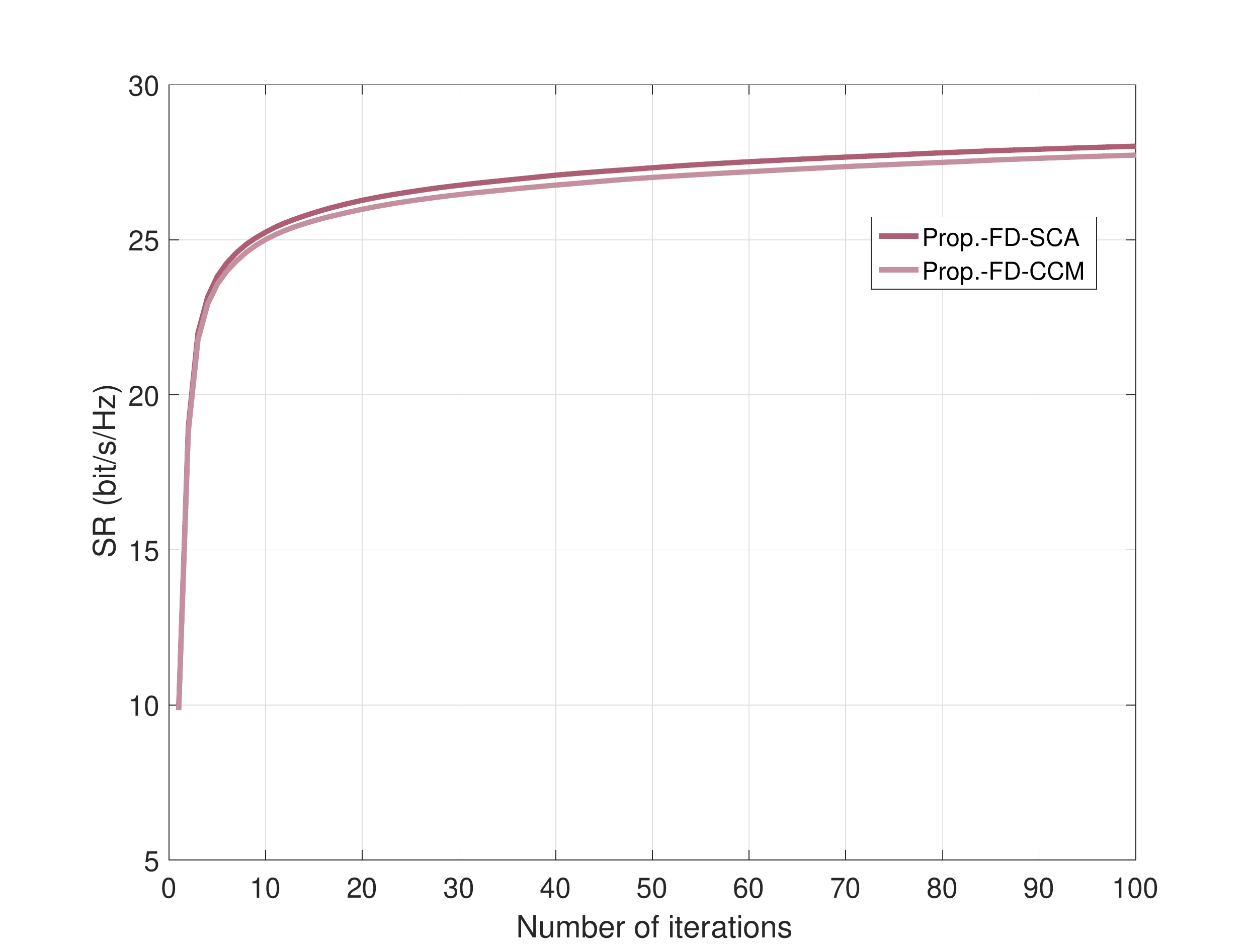}
	\caption{Convergence performance of different algorithms.}
	\label{fig:convergence}
\end{figure}

	\begin{figure}[t]
	\centering
	\includegraphics[scale=0.27]{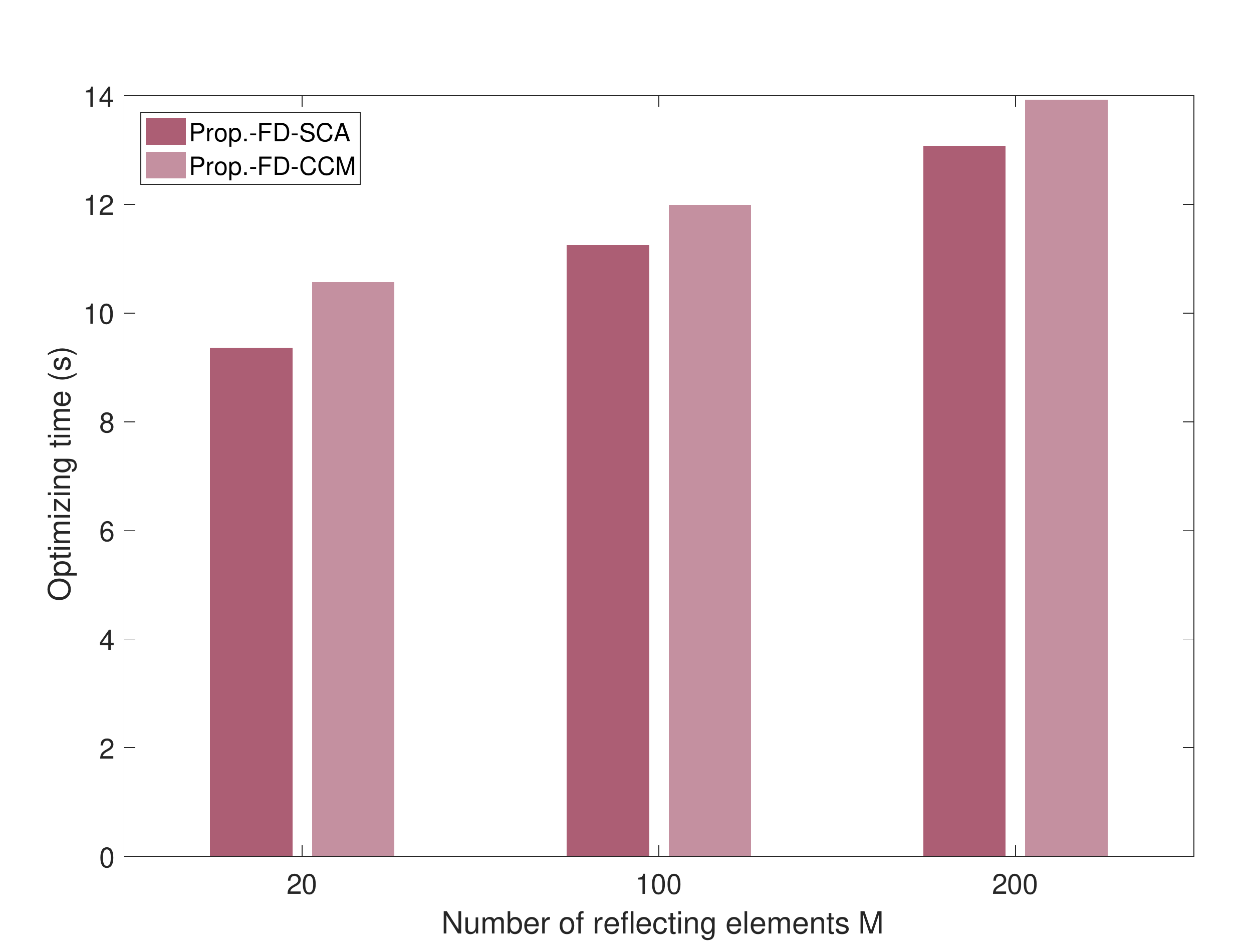}
	\caption{Optimization time comparison between different algorithms.}
	\label{fig:CPUtime}
\end{figure}

	\begin{figure}[t]
	\centering
	\includegraphics[scale=0.3]{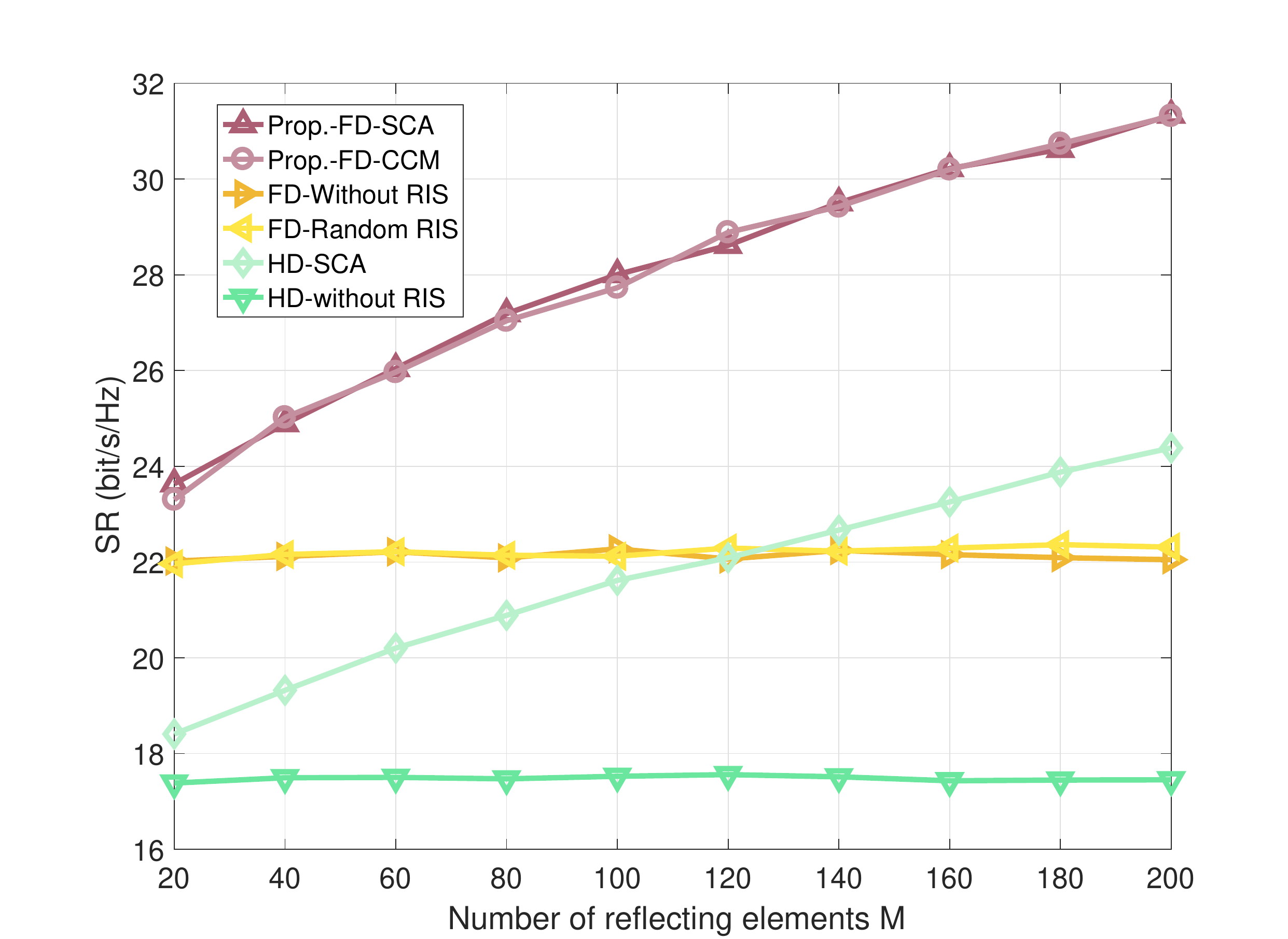}
	\caption{Achievable SR versus the number of reflecting elements $M$ under different deployment schemes.}
	\label{fig:DifferentElements}
\end{figure}		

\subsection{Simulation Results}
Firstly, we investigate the convergence of the proposed scheme to solve the original problem in (\ref{OF:original}), that is, Algorithm \ref{BCD-CCM} with CCM or SCA embedded. Figure \ref{fig:convergence} shows the SR versus the number of iterations under different algorithms. It can be observed that our proposed algorithms rise rapidly to a stable value within $20$ iterations, then converges within $100$ iterations. {Moreover, the SCA is slightly better than the CCM. To further see the difference between SCA and CCM, Fig. \ref{fig:CPUtime} compares the optimization time under different numbers of reflecting elements. It can be observed that for different size of RIS, the proposed algorithm with SCA requires less time than the CCM one. This is because the SCA scheme has a simpler optimization structure, as we have mentioned in Section IV-B.}

Figure \ref{fig:DifferentElements} compares the SR versus the number of reflecting elements $M$. It can be seen that our proposed schemes reach the highest SR. Besides, the scheme FD-Without RIS is much inferior to our proposed scheme.  For the FD with Random RIS scheme, it provides a much insignificant gain compared to that of FD-Without RIS, indicating that the potential gain of RIS could only be reaped by a well-designed phase configuration.  {Observing the curves of HD-SCA and Prop.-FD-SCA, we can see that with the increase of the reflecting elements, the SR of both keeps growing,  and the gap between them is enlarged as well.}  All these observations imply that configuring RIS to an FD networking system with designed phase shifts and TPC matrices can manage the interference environment effectively.

\begin{figure}[t]
	\centering
	\includegraphics[scale=0.3]{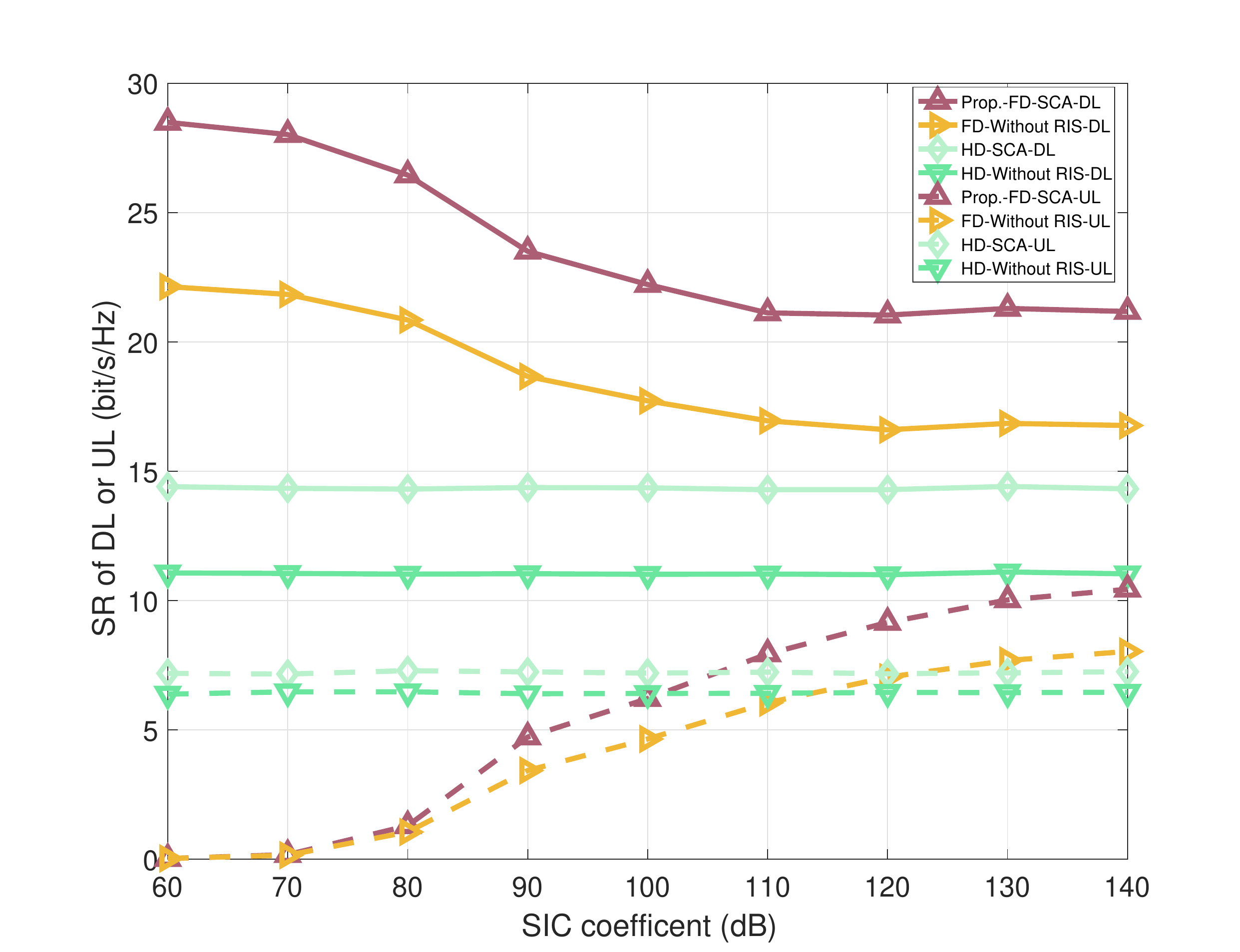}
	\caption{Achievable SR of DL or UL versus SIC coefficient ${\rho _{{_{l,l}}}}$.}
	\label{fig:DifferentSI}
\end{figure}

\begin{figure}[t]
	\centering
	\includegraphics[scale=0.3]{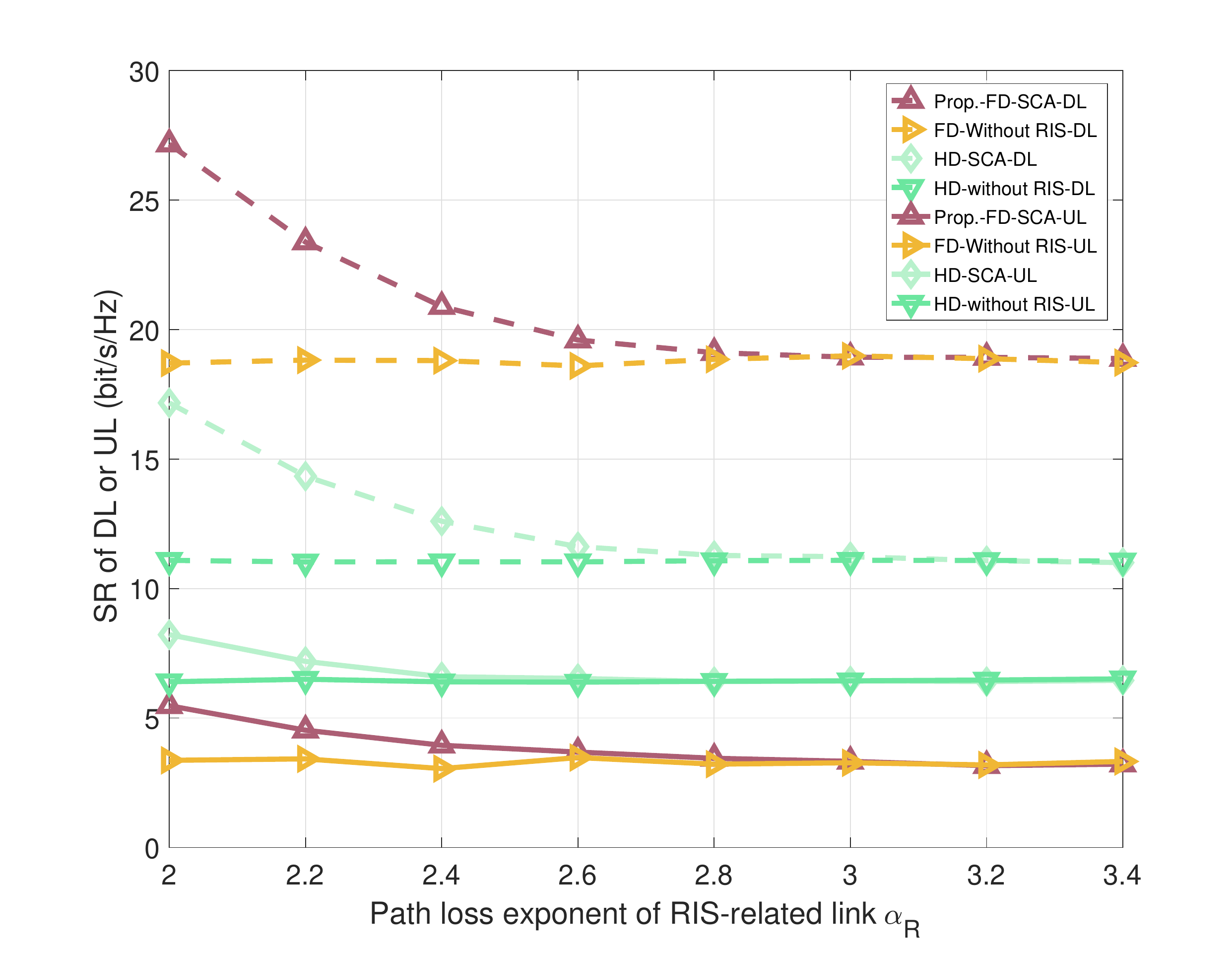}
	\caption{Achievable SR of DL or UL versus the path loss exponent of RIS-related links ${\alpha}_{{R}}$.}
	\label{fig:DifferentRISLoss}	
\end{figure}

\begin{figure}[t]
	\centering
	\includegraphics[scale=0.3]{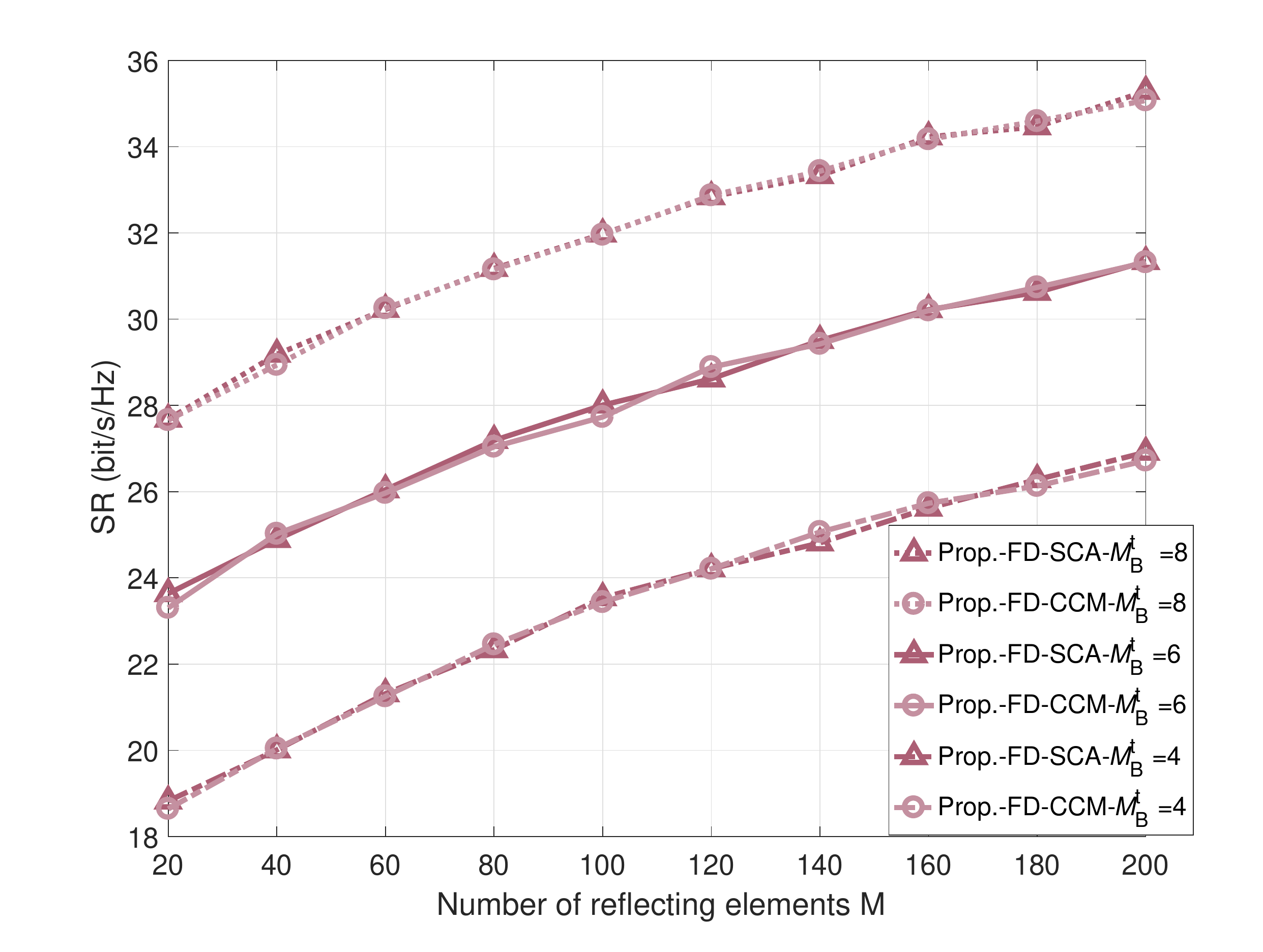}
	\vspace{0mm}
	\caption{Achievable SR versus the number of reflecting elements $M$ under different numbers of BS transmit antennas.}
	\label{fig:DifferentAntenna}
\end{figure}

	\begin{figure}[t]
	\centering
	\includegraphics[scale=0.3]{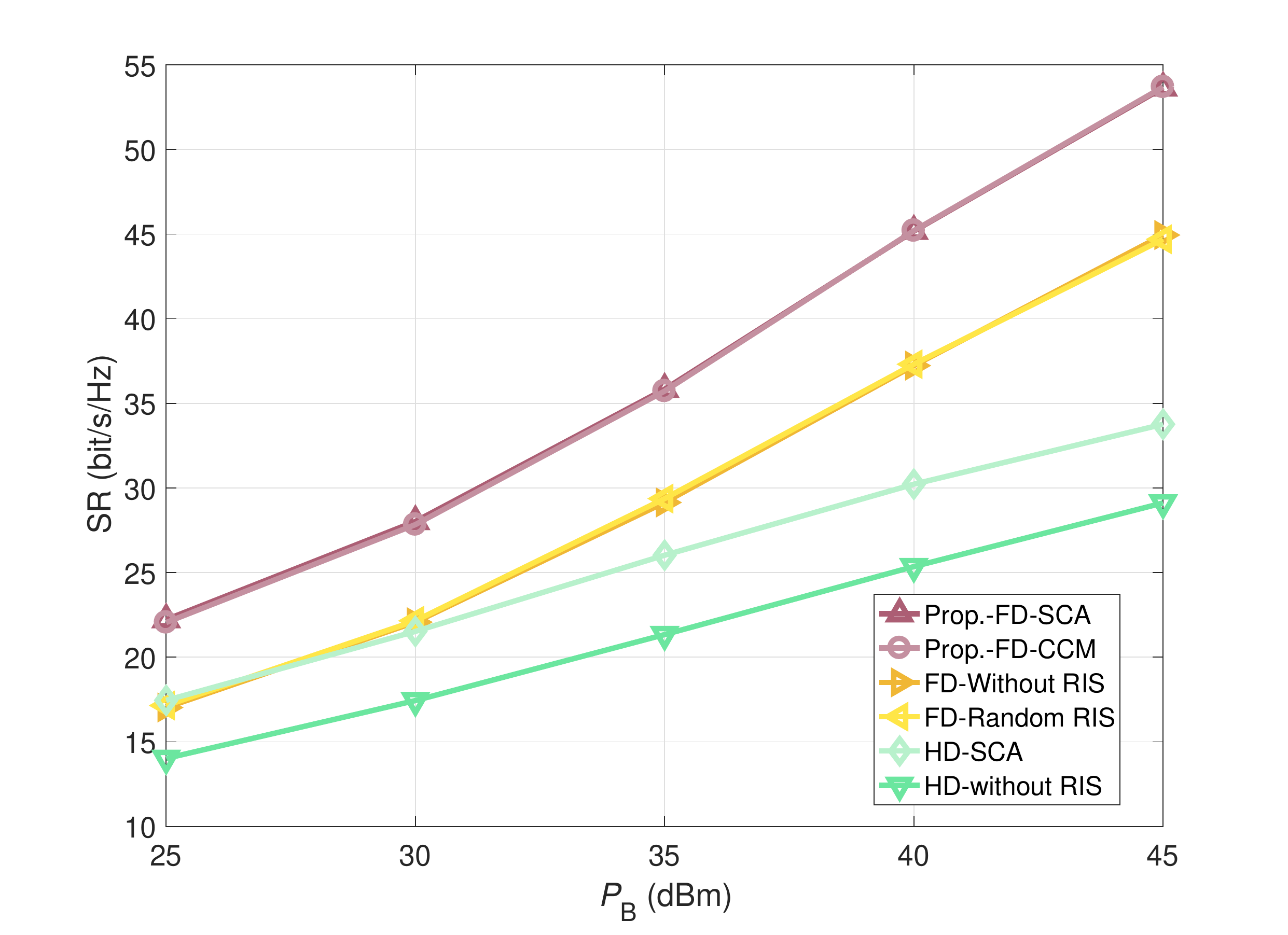}
	\vspace{5mm}
	\caption{Achievable SR versus BS transmit power $P_{B}$.}
	\label{fig:DifferentBSPower}
\end{figure}

	  In the following we study the impact of SIC capability on system performance. Figure \ref{fig:DifferentSI} depicts the SRs of downlink and uplink versus the SIC capability. For the SIC coefficient ${\rho _{{_{l,l}}}}$, a larger value in dB implies a more powerful SIC capability. It can be observed from Fig. \ref{fig:DifferentSI} that the uplink SR of all schemes except the HD ones raises rapidly with the increase of SIC coefficient. {Among them, for the FD scheme without RIS, even if the SIC coefficient reaches $140$ dB, the uplink performance gain is still limited, and is even inferior to that of the proposed scheme under $110$ dB.} This observation indicates that with jointly well-designed TPC and phase shift matrices, our proposed scheme can alleviate the requirement of SIC capability. Moreover, as the SIC coefficient increases, up to $130$ dB, both the UL and DL sum rate of the optimized scheme tend to be saturated. This is mainly because the power of the residual SI becomes insignificant among all interference perceived at the BS, and improving the SIC coefficient can no longer bring a significant gain. {We also notice that our proposed scheme outperforms the HD without RIS and  with optimized RIS when SIC is greater than $100$~dB and $105$~dB, respectively.} Above observation validates the superiority of our proposed scheme.
   
In all above simulations, RIS-related path loss exponent ${\alpha}_{{R}}$ is set as $2.2$, which is close to the path loss exponent in free space,  and may become impractical in some application scenarios. Therefore, we are motivated to study the influence of RIS-related path loss on the system performance. Figure \ref{fig:DifferentRISLoss} shows the SRs of downlink and uplink  versus the RIS-related path loss exponent ${\alpha}_{{R}}$. It can be observed that the SRs of both links decrease as the RIS-related path loss exponent increases. When ${\alpha}_R$ is greater than $2.8$, the SR gain obtained from RIS becomes pretty marginal. This is because the reflected link suffers from a double fading, leading to the result that the power strength of the RIS reflected link is much smaller than the direct one. Precisely, we find in our simulations that the path loss of the direct link from BS to the DL user center in Cell 1 is about $-123.19$ dB, while that of the reflected link is $-156.84$ dB and $-183.25$ dB when  ${\alpha _R}$ is set as $2.2$ and $2.8$, respectively. Obviously, with ${\alpha _R}=2.8,$ the deployment of RISs on the coordinate of $(350,0,15)$ makes nearly no sense. This provides an engineering insight for us, that is when we deploy the RIS in FD networks, it is better to choose the path loss of the reflected link as small as possible. Otherwise, RISs cannot bring significant performance gain, which is wasteful of hardware resources.

\begin{figure}[t]
	\centering
	\includegraphics[scale=0.3]{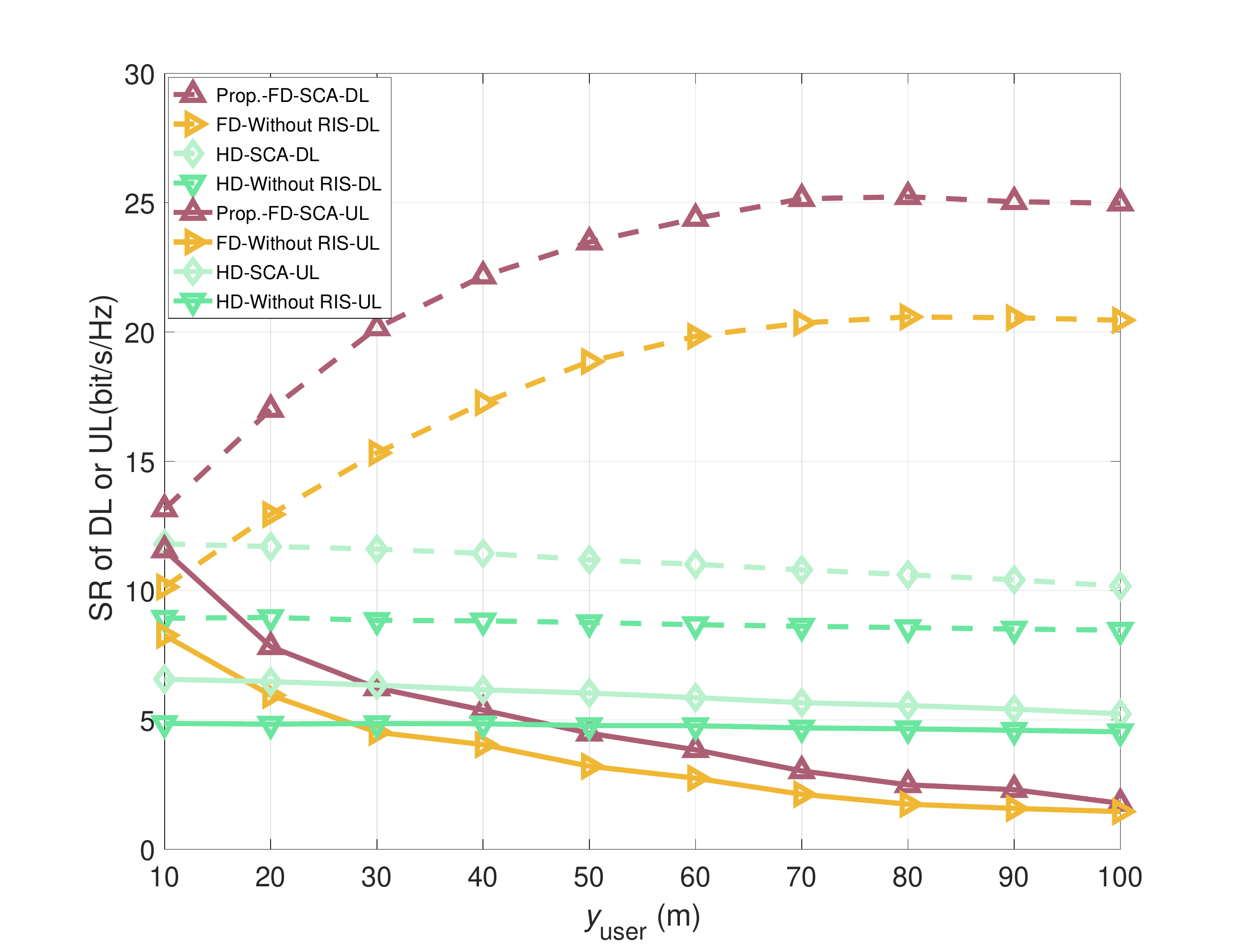}
	\caption{Achievable SR of DL or UL versus the distance from user center to the $y$-axis ${{{{y}}}_\text{{user}}}$.}
	\label{fig:DifferentYu}
\end{figure}	
	
\begin{figure}[t]
	\centering
	\includegraphics[scale=0.3]{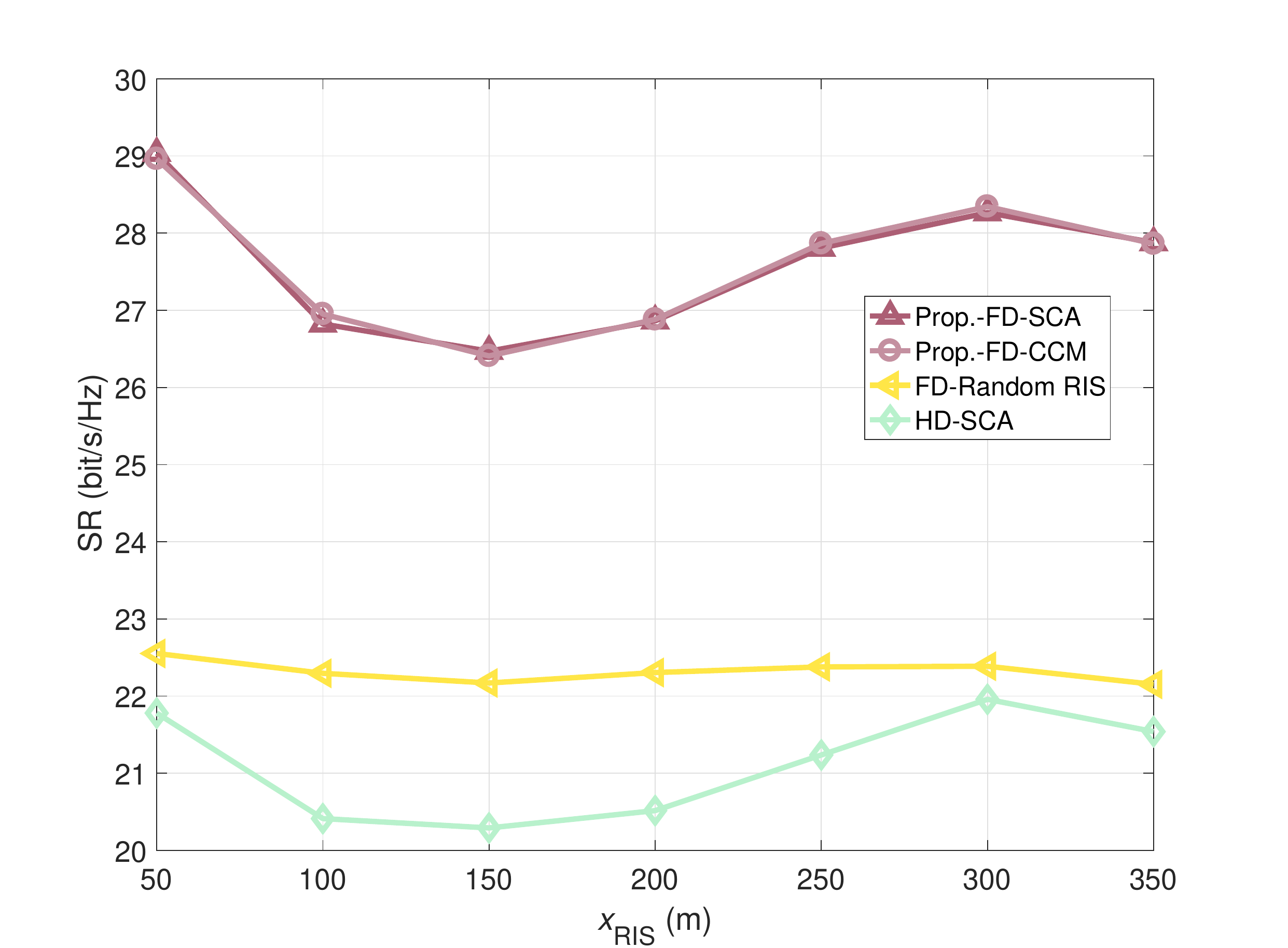}
	\caption{Achievable SR versus the $x$-axis location of  RIS ${{{x}}_\text{{RIS}}}$.}
	\label{fig:DifferentXr}
\end{figure}

\begin{figure}[t]
	\centering
	\includegraphics[scale=0.3]{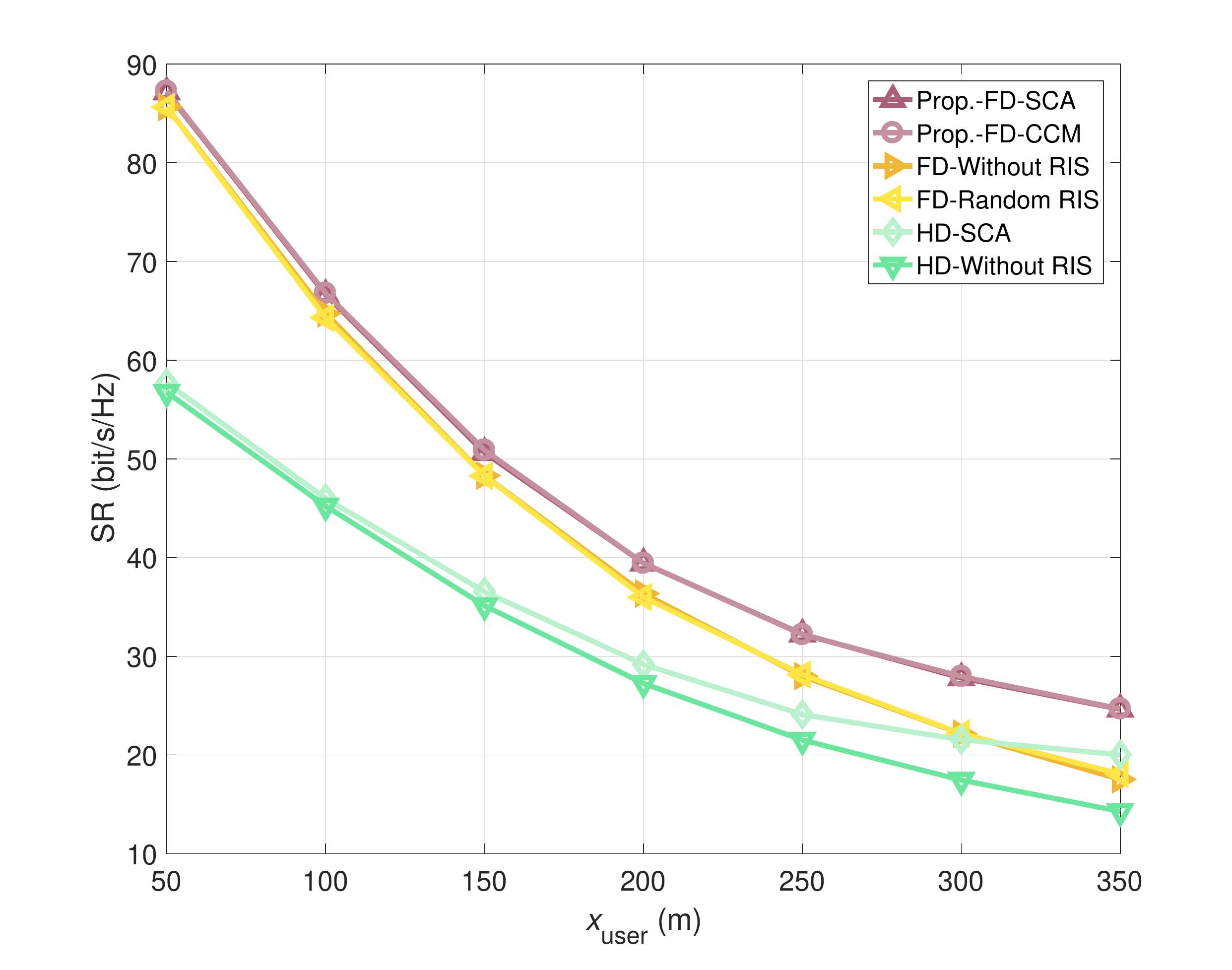}
	\caption{Achievable SR versus the $x$-axis distance from user center to cell center ${x_{\rm{user}}}$.}
	\label{fig:DifferentXu}
\end{figure}

Figure \ref{fig:DifferentAntenna} investigates the effect of different antenna configurations at BSs. As is known to all, more transmit antennas will bring more degrees of freedom, whilst the hardware cost also increases. Firstly, it can be observed from Fig. \ref{fig:DifferentAntenna} that the SR of the whole system increases as $\mathop M\nolimits_B^t $ becomes larger. Hence, it is promising to combine RIS with massive MIMO to further improve the system performance. From another perspective, increasing the size of antenna arrays at BSs requires more RF chains, which is much more costly than scaling up the RIS. Explicitly, comparing the curves of $\mathop M\nolimits_B^t  = 8$ and $\mathop M\nolimits_B^t  = 6{\rm{ }}$ in Fig. \ref{fig:DifferentAntenna}, with nearly $80$ more reflecting elements, the curve under  $\mathop M\nolimits_B^t  = 6$ reaches the SR as that under $\mathop M\nolimits_B^t  = 8$. {Hence RIS may serve as an cost-efficient way to improve the achievable performance, instead of upgrading whole BS transceivers with more RF chains.}

{Figure \ref{fig:DifferentBSPower} shows the SR versus different BS transmit power $\mathop P\nolimits_{B }$. It can be observed that as the transmit power increases, the SR of all schemes improves. Comparing the proposed scheme with the HD-SCA one, we find that the performance gain brought by RIS becomes larger as $\mathop P\nolimits_{B }$ increases. This is because when the transmit power at BSs becomes larger, more power resources are allocated to the DL users while the UL users suffer a partial loss. Thus, the increment and difference at SR between them are both enlarged. This observation proves that our proposed scheme is superior to the traditional ones.}	

{In the FD networking system, when UL and DL users are located closely to each other, the UE-UE interference is significant, deteriorating the receiving performance of DL users.} Here we study the effect of varying the distance between UL and DL users on the system performance. {In specific, the distance from the user center to the $y$-axis  is denoted as ${{{y}}_{\rm{user}}}$, and a larger value of ${{{y}}_{\rm{user}}}$ implies that UL and DL users in each cell are located more distantly. Figure \ref{fig:DifferentYu} shows the DL and UL SRs versus ${{{y}}_{\rm{user}}}$. Comparing  our proposed scheme with the one without RIS, the downlink SR performance has been improved effectively, which proves that RIS can well handle the UE-UE interference.} {Further, we can observe that when the ${{{y}}_{\rm{user}}}$ changes from 10 m to 70 m, the interference between users gradually decreases, and the SR of DL increases while that of UL decreases. There are two reasons for this, one is that the UE-UE interference perceived at the DL users is reduced, and the other is that  the UL users are far away from both BSs and RIS, which means that the path loss increase for both direct and reflected links. When ${{{y}}_{\rm{user}}}$ is larger than 70, the interference from UL to DL users is no longer dominant, so the SR of DL becomes saturated, while the SR of UL continues to decrease since the distance between UL users and BSs keeps increasing.}

Denote the coordinate of RIS in the $x$-axis as ${{{x}}_\text{{RIS}}}$. In Fig. \ref{fig:DifferentXr}, we study the impact of the RIS location by moving the RIS from ${{{x}}_\text{{RIS}}} = 50$ m to ${{{x}}_\text{{RIS}}} = 350{}$ m, which is from the center to the edge of cell 1. As shown in Fig. \ref{fig:DifferentXr}, all the schemes with RIS perform well when RIS is placed near the midpoint of two cells. Explicitly, the SR reaches its minimum and maximum at ${{{x}}_\text{{RIS}}}=150$ and ${{{x}}_\text{{RIS}}}=300$, respectively. Here we provide a theoretical analysis to these results. When the RIS moves in the first cell, it contributes more to the performance of cell 1. And the uplink and downlink users are symmetrically distributed, therefore, we mainly consider the path loss of the reflected link of DL users in cell 1. Denote ${{{d}}_{B,R}} = \sqrt {{{{x}}_\text{{RIS}}}^2  + {{\left( {30 - 15} \right)}^2}} $ and ${{{d}}_{R,U}} = \sqrt {\left( {{{{x}}_\text{{RIS}}} - 300} \right) + {{\left( {15 - 1.5} \right)}^2} + {{50}^2}}$ as the Euclidean distances between BS-RIS and RIS-user, respectively. Ignoring the small-scale fading, we express the large-scale channel gain of the reflected link from BS to users as $P{L_\text{{RIS}}} = 2{\rho _0} - 10{\alpha _R}{\log _{10}}\left( {{{{d}}_{B,R}}{{{d}}_{R,U}}} \right)$, which achieves its maximum and minimum at ${{{x}}_\text{{RIS}}}=158.7888$ m and ${{{x}}_\text{{RIS}}}=290.4796$ m, respectively. These analytical results agree with above observations. Though the SR reaches its maximum at ${{{x}}_\text{{RIS}}}=300$ m, the performance of cell 2 will suffer a certain loss. To achieve a performance balance in the practical system, RIS should be deployed in the middle of two cells.

{In Fig. \ref{fig:DifferentXu}, we study the impact of the $x$-axis distance from the user center to the severed BS, denoted as ${x_{\rm{user}}}$.
 As shown in Fig. \ref{fig:DifferentXu}, 
when users are distributed around the cell center, i.e., ${x_{\rm{user}}}$ is small, the gap between the proposed scheme and the FD with random RIS or without RIS is insignificant, indicating that RIS provides a limited regulatory effect even with optimized phase shifts. 
{When the users move forward the cell edge, for both FD and HD, all schemes with optimized RIS obtain a performance gain, compared to the counterparts without RIS and random RIS. Among these, our proposed scheme achieves the best cell-edge performance, validating that RIS deployment can promote FD networking performance at the edge coverage.}}

\begin{figure}[t]
	\centering
	\includegraphics[scale=0.3]{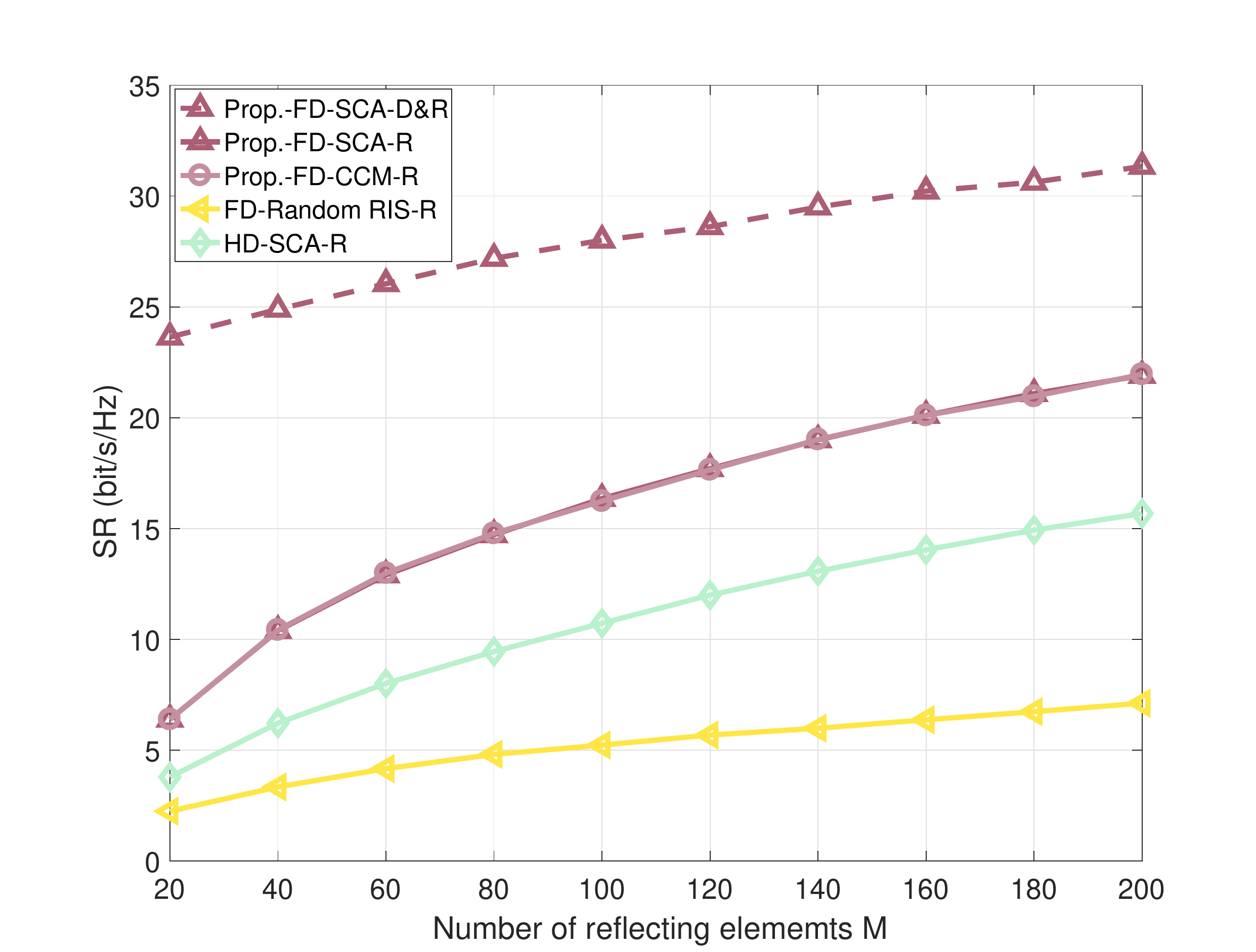}
	\caption{Achievable SR under the circumstance with or without direct links.}
	\label{fig:withoutdirectlink}
\end{figure}

Considering some ultra-dense communication scenarios, where direct links may be hindered by unintentional obstacles or malicious blockage, we investigate the system performance without direct links, that is when ${\alpha _{B,U}}$ tends to infinity, as shown in Fig. \ref{fig:withoutdirectlink}. In this figure, we use suffix D\&R or R to represent the scheme with or without direct links, respectively. Clearly, among all the schemes without direct links, our proposed design achieves the highest SR. This is because the RIS with optimized phase shifts provides a favorable reflected link that complements the absence of direct link. Moreover, comparing the D\&R  and R schemes both applied the proposed algorithm, when $M$ is equal to $100$, the latter scheme maintains the SE of $16$~bit/s/Hz, which reaches more than 50 percent of the former one. This proportion will increase when scaling up the RIS. This observation hints that, when we deploy an RIS with enough reflecting elements and a well-designed configuration, it can guarantee a decent experience of communication rates even in the scenario without direct links. 

		\section{Conclusions}
		In this paper, we have proposed a multi-cell FD system for next-generation mobile networking by deploying RISs at cell boundaries. To achieve the full potential of such deployment, the SR maximization problem has been formulated to jointly optimize the TPC matrices at BSs and users and phase shifts at RIS under the power and unit modulus constraints. To solve this non-convex problem, we have decoupled it into a pair of subproblems by exploiting the relationship between SR and MMSE. The optimal TPC matrices have been obtained in closed form, while the phase shift matrix has been designed to reach the suboptimal solution through a pair of efficient algorithms, i.e., SCA and CCM. {Simulation results have shown that the SCA algorithm has lower complexity, and the proposed scheme outperforms other traditional FD and HD networking benchmarks.} Besides, we have validated that the RIS deployment could reduce the requirements for SIC and the number of BS antennas in FD networks, which further reduces the power consumption and hardware cost. {Moreover, applying the RIS with enough reflecting elements and well-designed configurations can alleviate the UE-UE interference especially at the cell edge, and guarantee a decent experience in some blockage scenarios.}
		In a nutshell, the utilization of RIS has made the FD networking system become efficient and practical.	

		
		
	    \balance 
		\bibliographystyle{IEEEtran}
		\bibliography{ref}

\begin{thebibliography}{10}
\providecommand{\url}[1]{#1}
\csname url@samestyle\endcsname
\providecommand{\newblock}{\relax}
\providecommand{\bibinfo}[2]{#2}
\providecommand{\BIBentrySTDinterwordspacing}{\spaceskip=0pt\relax}
\providecommand{\BIBentryALTinterwordstretchfactor}{4}
\providecommand{\BIBentryALTinterwordspacing}{\spaceskip=\fontdimen2\font plus
\BIBentryALTinterwordstretchfactor\fontdimen3\font minus
  \fontdimen4\font\relax}
\providecommand{\BIBforeignlanguage}[2]{{%
\expandafter\ifx\csname l@#1\endcsname\relax
\typeout{** WARNING: IEEEtran.bst: No hyphenation pattern has been}%
\typeout{** loaded for the language `#1'. Using the pattern for}%
\typeout{** the default language instead.}%
\else
\language=\csname l@#1\endcsname
\fi
#2}}
\providecommand{\BIBdecl}{\relax}
\BIBdecl

\bibitem{FullDuplexMultiUserMultiCell}
P.~Aquilina, A.~C. Cirik, and T.~Ratnarajah, ``Weighted sum rate maximization
  in full-duplex multi-user multi-cell {MIMO} networks,'' \emph{IEEE
  Transactions on Communications}, vol.~65, no.~4, pp. 1590--1608, Mar. 2017.

\bibitem{FullDuplexWirelessCommunications}
Z.~Zhang, K.~Long, A.~V. Vasilakos, and L.~Hanzo, ``Full-duplex wireless
  communications: Challenges, solutions, and future research directions,''
  \emph{Proceedings of the IEEE}, vol. 104, no.~7, pp. 1369--1409, July 2016.

\bibitem{DynamicSpectrumSharing}
S.~K. Sharma, T.~E. Bogale, L.~B. Le, S.~Chatzinotas, X.~Wang, and
  B.~Ottersten, ``Dynamic spectrum sharing in {5G} wireless networks with
  full-duplex technology: Recent advances and research challenges,'' \emph{IEEE
  Communications Surveys \& Tutorials}, vol.~20, no.~1, pp. 674--707, 2018.

\bibitem{Fullduplextechniquesfor5G}
Z.~Zhang, X.~Chai, K.~Long, A.~V. Vasilakos, and L.~Hanzo, ``Full duplex
  techniques for {5G} networks: self-interference cancellation, protocol
  design, and relay selection,'' \emph{IEEE Communications Magazine}, vol.~53,
  no.~5, pp. 128--137, May 2015.

\bibitem{Applicationsofselfinterferencecancellation}
S.~Hong, J.~Brand, J.~I. Choi, M.~Jain, J.~Mehlman, S.~Katti, and P.~Levis,
  ``Applications of self-interference cancellation in {5G} and beyond,''
  \emph{IEEE Communications Magazine}, vol.~52, no.~2, pp. 114--121, Feb. 2014.

\bibitem{Fullduplexdevicetodevice}
S.~Ali, N.~Rajatheva, and M.~Latva-aho, ``Full duplex device-to-device
  communication in cellular networks,'' in \emph{Proceedings of European
  Conference on Networks and Communications (EuCNC)}, 2014, pp. 1--5.

\bibitem{fullduplexsmallcellwireless}
D.~Nguyen, L.-N. Tran, P.~Pirinen, and M.~Latva-aho, ``On the spectral
  efficiency of full-duplex small cell wireless systems,'' \emph{IEEE
  Transactions on Wireless Communications}, vol.~13, no.~9, pp. 4896--4910,
  Sept. 2014.

\bibitem{FDcellualr}
S.~Goyal, P.~Liu, S.~S. Panwar, R.~A. Difazio, R.~Yang, and E.~Bala, ``Full
  duplex cellular systems: will doubling interference prevent doubling
  capacity?'' \emph{IEEE Communications Magazine}, vol.~53, no.~5, pp.
  121--127, May 2015.

\bibitem{FD_ActiveBeamforming1}
J.~Bai and A.~Sabharwal, ``Asymptotic analysis of {MIMO} multi-cell full-duplex
  networks,'' \emph{IEEE Transactions on Wireless Communications}, vol.~16,
  no.~4, pp. 2168--2180, April 2017.

\bibitem{Smartradioenvironments}
M.~Di~Renzo, A.~Zappone, M.~Debbah, M.-S. Alouini, C.~Yuen, J.~de~Rosny, and
  S.~Tretyakov, ``Smart radio environments empowered by reconfigurable
  intelligent surfaces: How it works, state of research, and the road ahead,''
  \emph{IEEE Journal on Selected Areas in Communications}, vol.~38, no.~11, pp.
  2450--2525, Nov. 2020.

\bibitem{MultiuserFullDuplexTwoWayCommunications}
Z.~Peng, Z.~Zhang, C.~Pan, L.~Li, and A.~L. Swindlehurst, ``Multiuser
  full-duplex two-way communications via intelligent reflecting surface,''
  \emph{IEEE Transactions on Signal Processing}, vol.~69, pp. 837--851, 2021.

\bibitem{UAVCommunicationWithRIS}
K.~Tian, B.~Duo, S.~Li, Y.~Zuo, and X.~Yuan, ``Hybrid uplink and downlink
  transmissions for full-duplex {UAV} communication with {RIS},'' \emph{IEEE
  Wireless Communications Letters}, vol.~11, no.~4, pp. 866--870, April 2022.

\bibitem{IntegratedSensingandCommunication}
X.~Wang, Z.~Fei, J.~Huang, and H.~Yu, ``Joint waveform and discrete phase shift
  design for {RIS}-assisted integrated sensing and communication system under
  cramer-rao bound constraint,'' \emph{IEEE Transactions on Vehicular
  Technology}, vol.~71, no.~1, pp. 1004--1009, Jan. 2022.

\bibitem{ChenTWC}
Y.~Chen, M.~Wen, E.~Basar, Y.-C. Wu, L.~Wang, and W.~Liu, ``Exploiting
  reconfigurable intelligent surfaces in edge caching: Joint hybrid beamforming
  and content placement optimization,'' \emph{IEEE Transactions on Wireless
  Communications}, vol.~20, no.~12, pp. 7799--7812, Dec. 2021.

\bibitem{IntegratedAccessandBackhaul}
G.~Y. Suk, S.-M. Kim, J.~Kwak, S.~Hur, E.~Kim, and C.-B. Chae, ``Full duplex
  integrated access and backhaul for {5G} nr: Analyses and prototype
  measurements,'' \emph{IEEE Wireless Communications}, vol.~29, no.~4, pp.
  40--46, Aug. 2022.

\bibitem{DesignAndAnalysis}
J.~Zhang, H.~Luo, N.~Garg, A.~Bishnu, M.~Holm, and T.~Ratnarajah, ``Design and
  analysis of wideband in-band-full- duplex {FR2-IAB} networks,'' \emph{IEEE
  Transactions on Wireless Communications}, vol.~21, no.~6, pp. 4183--4196,
  Jun. 2022.

\bibitem{JointAnalog}
M.~A. Islam, G.~C. Alexandropoulos, and B.~Smida, ``Joint analog and digital
  transceiver design for wideband full duplex {MIMO} systems,'' \emph{IEEE
  Transactions on Wireless Communications}, \emph{to be published,} 2022.

\bibitem{HybridBeamforming}
I.~P. Roberts, J.~G. Andrews, and S.~Vishwanath, ``Hybrid beamforming for
  millimeter wave full-duplex under limited receive dynamic range,'' \emph{IEEE
  Transactions on Wireless Communications}, vol.~20, no.~12, pp. 7758--7772,
  Dec. 2021.

\bibitem{MillimeterWave}
L.~Zhu, J.~Zhang, Z.~Xiao, X.~Cao, X.-G. Xia, and R.~Schober, ``Millimeter-wave
  full-duplex {UAV} relay: Joint positioning, beamforming, and power control,''
  \emph{IEEE Journal on Selected Areas in Communications}, vol.~38, no.~9, pp.
  2057--2073, Sept. 2020.

\bibitem{Chen20TNSE}
Y.~Chen, L.~Wang, R.~Ma, W.~Liu, M.~Wen, and A.~Fei, ``Performance analysis of
  heterogeneous networks with wireless caching and full duplex relaying,''
  \emph{IEEE Transactions on Network Science and Engineering}, vol.~7, no.~4,
  pp. 2429--2442, 1 Oct.-Dec. 2020.

\bibitem{MaximizingSecondary}
R.~Jafari and A.~O. Fapojuwo, ``Maximizing secondary users’ sum-throughput in
  an in-band full-duplex cognitive wireless powered backscatter communication
  network,'' \emph{IEEE Systems Journal}, vol.~16, no.~3, pp. 4082--4093, Sept.
  2022.

\bibitem{MultiDomain}
J.~Hu, Y.~Zheng, and K.~Yang, ``Multi-domain resource scheduling for
  full-duplex aided wireless powered communication network,'' \emph{IEEE
  Transactions on Vehicular Technology}, vol.~71, no.~10, pp. 10\,849--10\,862,
  Oct. 2022.

\bibitem{WaveformDesign}
Z.~Xiao and Y.~Zeng, ``Waveform design and performance analysis for full-duplex
  integrated sensing and communication,'' \emph{IEEE Journal on Selected Areas
  in Communications}, vol.~40, no.~6, pp. 1823--1837, Jun. 2022.

\bibitem{FullDuplexMIMOInterferenceChannels}
A.~C. Cirik, R.~Wang, Y.~Hua, and M.~Latva-aho, ``Weighted sum-rate
  maximization for full-duplex {MIMO} interference channels,'' \emph{IEEE
  Transactions on Communications}, vol.~63, no.~3, pp. 801--815, March 2015.

\bibitem{fullduplexmassiveMIMOsystems}
X.~Wang, D.~Zhang, K.~Xu, and C.~Yuan, ``On the sum rate of multi-user
  full-duplex massive {MIMO} systems,'' in \emph{Proceedings of IEEE
  International Conference on Communication Systems (ICCS)}, 2016, pp. 1--7.

\bibitem{LowComplexity}
M.~Zou, J.~Ma, and B.~Jiao, ``Low-complexity coordinated beamforming for full
  duplex {MIMO} wireless cellulars,'' \emph{IEEE Access}, vol.~6, pp.
  33\,225--33\,237, Jun. 2018.

\bibitem{LinearTransceiverDesignforFullDuplex}
A.~C. Cirik, O.~Taghizadeh, L.~Lampe, R.~Mathar, and Y.~Hua, ``Linear
  transceiver design for full-duplex multi-cell {MIMO} systems,'' \emph{IEEE
  Access}, vol.~4, pp. 4678--4689, 2016.

\bibitem{Ming_min_Zhao}
M.-M. Zhao, Y.~Cai, M.-J. Zhao, Y.~Xu, and L.~Hanzo, ``Robust joint hybrid
  analog-digital transceiver design for full-duplex {mmWave} multicell
  systems,'' \emph{IEEE Transactions on Communications}, vol.~68, no.~8, pp.
  4788--4802, Aug. 2020.

\bibitem{Wang_FD_Hybrid}
G.~Wang, Z.~Yang, and T.~Gong, ``A two-stage hybrid beamforming design for
  full-duplex {mmWave} communications,'' in \emph{Proceedings of International
  Wireless Communications and Mobile Computing (IWCMC)}, 2022, pp. 790--795.

\bibitem{Fullduplexcellularsystems}
S.~Goyal, P.~Liu, S.~S. Panwar, R.~A. Difazio, R.~Yang, and E.~Bala, ``Full
  duplex cellular systems: will doubling interference prevent doubling
  capacity?'' \emph{IEEE Communications Magazine}, vol.~53, no.~5, pp.
  121--127, May 2015.

\bibitem{CascadedChannel}
S.~Lin, M.~Wen, and F.~Chen, ``Cascaded channel estimation using full duplex
  for {IRS}-aided multiuser communications,'' in \emph{Proceedings of IEEE
  Wireless Communications and Networking Conference (WCNC)}, 2022, pp.
  375--380.

\bibitem{DeepLearning}
S.~Zhang, S.~Zhang, F.~Gao, J.~Ma, and O.~A. Dobre, ``Deep learning-based {RIS}
  channel extrapolation with element-grouping,'' \emph{IEEE Wireless
  Communications Letters}, vol.~10, no.~12, pp. 2644--2648, Dec. 2021.

\bibitem{RISAidedWirelessCommunication}
Y.~Cheng, W.~Peng, C.~Huang, G.~C. Alexandropoulos, C.~Yuen, and M.~Debbah,
  ``Ris-aided wireless communications: Extra degrees of freedom via rotation
  and location optimization,'' \emph{IEEE Transactions on Wireless
  Communications}, vol.~21, no.~8, pp. 6656--6671, Aug. 2022.

\bibitem{WirelessCoverageExtension}
S.~Zeng, H.~Zhang, B.~Di, Z.~Han, and L.~Song, ``Reconfigurable intelligent
  surface ({RIS}) assisted wireless coverage extension: {RIS} orientation and
  location optimization,'' \emph{IEEE Communications Letters}, vol.~25, no.~1,
  pp. 269--273, Jan. 2021.

\bibitem{ActiveReconfigurableIntelligentSurface}
R.~Long, Y.-C. Liang, Y.~Pei, and E.~G. Larsson, ``Active reconfigurable
  intelligent surface-aided wireless communications,'' \emph{IEEE Transactions
  on Wireless Communications}, vol.~20, no.~8, pp. 4962--4975, Aug. 2021.

\bibitem{MultipleReconfigurableIntelligentSurfaces}
B.~C. Nguyen, T.~M. Hoang, P.~T. Tran, T.~N. Nguyen, V.-D. Phan, B.~V. Minh,
  and M.~Voznak, ``Cooperative communications for improving the performance of
  bidirectional full-duplex system with multiple reconfigurable intelligent
  surfaces,'' \emph{IEEE Access}, vol.~9, pp. 134\,733--134\,742, 2021.

\bibitem{InterceptProbability}
O.~Zaghdoud, A.~B. Mnaouer, and H.~Boujemaa, ``Intercept probability and
  secrecy capacity analysis of {RIS}-based wireless communication with
  full-duplex receiver,'' in \emph{Proceedings of International Wireless
  Communications and Mobile Computing (IWCMC)}, 2021, pp. 843--848.

\bibitem{aReconfigurableIntelligentSurfaceandaRelay}
M.~Obeed and A.~Chaaban, ``Joint beamforming design for multiuser {MISO}
  downlink aided by a reconfigurable intelligent surface and a relay,''
  \emph{IEEE Transactions on Wireless Communications}, vol.~21, no.~10, pp.
  8216--8229, Oct. 2022.

\bibitem{RobustBeamforming}
H.~Gao, K.~Cui, C.~Huang, and C.~Yuen, ``Robust beamforming for {RIS}-assisted
  wireless communications with discrete phase shifts,'' \emph{IEEE Wireless
  Communications Letters}, vol.~10, no.~12, pp. 2619--2623, Oct. 2021.

\bibitem{HardwareImpairments}
G.~Zhou, C.~Pan, H.~Ren, K.~Wang, and Z.~Peng, ``Secure wireless communication
  in {RIS}-aided {MISO} system with hardware impairments,'' \emph{IEEE Wireless
  Communications Letters}, vol.~10, no.~6, pp. 1309--1313, Oct. 2021.

\bibitem{OptimalControlforFullDuplex}
Z.~Yang, C.~Huang, J.~Shi, Y.~Chau, W.~Xu, Z.~Zhang, and M.~Shikh-Bahaei,
  ``Optimal control for full-duplex communications with reconfigurable
  intelligent surface,'' in \emph{Proceedings of IEEE International Conference
  on Communications (ICC)}, 2021, pp. 1--6.

\bibitem{SumRateOptimization}
Y.~Zhang, C.~Zhong, Z.~Zhang, and W.~Lu, ``Sum rate optimization for two way
  communications with intelligent reflecting surface,'' \emph{IEEE
  Communications Letters}, vol.~24, no.~5, pp. 1090--1094, May 2020.

\bibitem{WeightedSumRateMaximization}
H.~Guo, Y.-C. Liang, J.~Chen, and E.~G. Larsson, ``Weighted sum-rate
  maximization for reconfigurable intelligent surface aided wireless
  networks,'' \emph{IEEE Transactions on Wireless Communications}, vol.~19,
  no.~5, pp. 3064--3076, May 2020.

\bibitem{MulticellMIMOCommunications}
C.~Pan, H.~Ren, K.~Wang, W.~Xu, M.~Elkashlan, A.~Nallanathan, and L.~Hanzo,
  ``Multicell {MIMO} communications relying on intelligent reflecting
  surfaces,'' \emph{IEEE Transactions on Wireless Communications}, vol.~19,
  no.~8, pp. 5218--5233, Aug. 2020.

\bibitem{WeightedMMSEApproach}
Q.~Shi, M.~Razaviyayn, Z.-Q. Luo, and C.~He, ``An iteratively weighted {MMSE}
  approach to distributed sum-utility maximization for a {MIMO} interfering
  broadcast channel,'' \emph{IEEE Transactions on Signal Processing}, vol.~59,
  no.~9, pp. 4331--4340, Sept. 2011.

\bibitem{zhang2017matrix}
X.~Zhang, \emph{Matrix analysis and applications}.\hskip 1em plus 0.5em minus
  0.4em\relax Cambridge University Press, 2017.

\bibitem{Aunifiedconvergenceanalysis}
M.~Razaviyayn, M.~Hong, and Z.-Q. Luo, ``A unified convergence analysis of
  block successive minimization methods for nonsmooth optimization,''
  \emph{SIAM Journal on Optimization}, vol.~23, no.~2, pp. 1126--1153, 2013.

\end{thebibliography}

	\end{document}